# A GENERALISED MATCHING DISTRIBUTION FOR THE PROBLEM OF COINCIDENCES

B. O'NEILL,[*] *Australian National University*[**]

WRITTEN 19 NOVEMBER 2021

**Abstract**

This paper examines the classical matching distribution arising in the "problem of coincidences". We generalise the classical matching distribution with a preliminary round of allocation where items are correctly matched with some fixed probability, and remaining non-matched items are allocated using simple random sampling without replacement. Our generalised matching distribution is a convolution of the classical matching distribution and the binomial distribution. We examine the properties of this latter distribution and show how its probability functions can be computes. We also show how to use the distribution for matching tests and inferences of matching ability.

PROBLEM OF COINCIDENCES; HAT-CHECK PROBLEM; SECRET-SANTA PROBLEM; MATCHING STATISTIC; CLASSICAL MATCHING DISTRIBUTION; GENERALISED MATCHING DISTRIBUTION; MATCHING TEST

In this paper we examine a probability distribution arising in the "problem of coincidences", an antique probability problem dating back to Mortmont (1708, 1713). The problem arises in the classical "game of thirteen" (*jeu du treize*), where a dealer shuffles thirteen playing cards of a single suit and then turns them over in order; if no card appears in its proper place in the order (e.g., ascending order from ace, two, …, jack, queen, king), the dealer takes the stakes, but if there are any cards appearing in their proper place, the dealer pays each of the players. Another variation on this problem is to shuffle two groups of playing cards with the same numbers in different suits, and reveal one card in each group at a time, seeing if there is a "match" between the numbers shown on the two cards.

There are many names for this probability problem, corresponding to many equivalent forms. Penrice (1991) notes that "[t]here seem to be as many ways of describing the problem as there are names for it, and each description suggests some generalization…" (p. 617). One form of the problem, called the "hat-check problem", stipulates that $n$ gentlemen check-in their hats at a restaurant, but the hat-check girl accidentally mixes them up, and returns the hats at random. Another variation, called the "Secret-Santa problem" considers a group of $n$ people who draw the names in the group at random to determine who will buy a present for whom. In both cases we are interested in the distribution of the number of "matches" based on a fixed place for each item. The mathematics of the problem of coincidences has been considered by a number of authors and has been of recurring interest (e.g., Anderson 1943; Feller 1968; Abramson 1973;

---

[*] E-mail address: ben.oneill@hotmail.com.
[**] Research School of Population Health, Australian National University, Canberra ACT 0200, Australia.

Scoville 1980; Penrice 1991; Crain 1992, Knudsen and Skau 1996). The history of the problem is examined in Takács (1980), and unsurprisingly, it is tied up with the historical derivation of the subfactorial numbers (representing the number of permutations with no matches).

## 1. The classical matching distribution

Each variant of the problem can be formulated in general terms by taking a set of $n$ distinct items (labelled here as consecutive integers). Suppose we permute the order of these items at random to obtain a permutation vector $\mathbf{x} = (x_1, \ldots, x_n)$, assumed to be equiprobable over all permutations of $1, \ldots, n$ (i.e., simple random sampling without replacement). We then define the **matching statistic** by the number of "fixed points" in the permutation vector:

$$K = \sum_{i=1}^{n} \mathbb{I}(x_i = i).$$

The matching statistic counts the number of items in the original vector $\mathbf{s}$ that are in the same place after a random permutation. The classical version of the problem focused specifically on the probability that there are no matches, which is $\mathbb{P}(K = 0)$. However, it is simple to extend the problem to consider the full distribution of the number of matches, which fully describes the stochastic behaviour of the matching statistic. This distribution (which we will call the "classical matching distribution") is defined below.

**DEFINITION (Classical matching distribution):** This is a family of discrete distributions characterised by the probability mass function:[1]

$$\text{Match}(k|n) = \frac{1}{k!} \sum_{i=0}^{n-k} \frac{(-1)^i}{i!} \qquad k = 0, \ldots, n,$$

where we have a **size parameter** $n = 0, 1, 2, \ldots, \infty$. In the special case where $n = \infty$ we take $\text{Match}(k|\infty) \equiv \text{Pois}(k|1)$ by convention (based on limiting results shown below). □

We begin by demonstrating that the matching statistic follows the matching distribution. Since each permutation of the initial items is (by assumption) equiprobable, there are $n!$ equiprobable outcomes for the permutation vector $\mathbf{x}$. The number of outcomes with $K = k$ can be obtained by choosing $k$ fixed values from the $n$ elements and then counting the number of derangements for the remaining $n - k$ elements, which we denote as $\mathcal{D}(n - k)$. The number of derangements is given by the subfactorial numbers, whose properties have been extensively studied (see e.g., Hassani 2003 and Hassani 2004). Applying the fundamental principle of counting then gives:

$$|\{\mathbf{x} | K = k\}| = \binom{n}{k} \times \mathcal{D}(n - k).$$

---

[1] In the special case where $n = 0$ the distribution is a point-mass on $k = 0$, and in the special case where $n = 1$ the distribution is a point-mass on $k = 1$.

Consequently, under the assumption that all possible permutations are equiprobable, we have:

$$\mathbb{P}(K = k) = \frac{|\{\mathbf{x} | K = k\}|}{|\{\mathbf{x}\}|}$$

$$= \frac{1}{n!} \times \binom{n}{k} \times \mathcal{D}(n - k)$$

$$= \frac{1}{k!} \cdot \frac{\mathcal{D}(n - k)}{(n - k)!}.$$

The values $\mathcal{D}(n - k)$ are the **subfactorials**, defined recursively by the base values $\mathcal{D}(0) = 1$ and $\mathcal{D}(1) = 0$ and the recursion $\mathcal{D}(n - k) = (n - k - 1)[\mathcal{D}(n - k - 1) + \mathcal{D}(n - k - 2)]$. They can also be written in explicit form (in terms of the factorials) as:

$$\mathcal{D}(n - k) = (n - k)! \sum_{i=0}^{n-k} \frac{(-1)^i}{i!}.$$

Consequently, we can write the probabilities for the matching statistic as:

$$\mathbb{P}(K = k) = \frac{1}{k!} \sum_{i=0}^{n-k} \frac{(-1)^i}{i!} = \text{Match}(k|n).$$

In the definition of the matching distribution, we have allowed the special case $n = \infty$, which we conceive as a limiting case. To obtain this special case we examine the limiting form of the distribution that occurs when $n \to \infty$. In this case, using a well-known limit result for the ratio of derangements over permutations (see e.g., Hassani 2003, pp. 1-2) we get:

$$\lim_{n \to \infty} \mathbb{P}(K = k) = \frac{1}{k!} \lim_{n \to \infty} \frac{\mathcal{D}(n - k)}{(n - k)!} = \frac{1}{k!} \cdot \frac{1}{e} = \text{Pois}(k|1).$$

We can see that the distribution approaches the Poisson distribution with unit scale in the limit as $n \to \infty$. This result has been noted in much of the literature on the problem of coincidences (see e.g., Penrice 1991, Crain 1992, Knudsen and Skau 1996), and it implies the asymptotic probability $\lim_{n \to \infty} \mathbb{P}(K = 0) = 1/e$ which gives an approximate solution to the original problem. In our definition of the matching distribution we have allowed $n = \infty$ as an explicit case, and we set the distribution equal to its limit in this case. This convention has the advantage of giving a family of distributions that is closed under limits on its size parameter.

**REMARK:** One noteworthy aspect of the matching statistic is that we cannot have $K = n - 1$, and therefore the matching distribution always has $\mathbb{P}(K = n - 1) = 0$. Intuitively, this reflects the fact that if $n - 1$ items are matched in the permutation then the last item must also be matched —i.e., it is not possible to match all but one item in any permutation. Consequently, the matching distribution has a strange looking support, with a missing value at $n - 1$. Since the distribution has a non-zero probability at $K = n$ this also means that — except in the trivial cases $n = 0,1$ where the distribution is a point-mass— the distribution is not quasi-concave (unimodal) and is instead slightly bimodal. This slight bimodality of the distribution may need to be taken into account in some cases where the user forms a highest density region (HDR) for the distribution, though it is marginal even in this case. ◻

## 2. Important properties of the classical matching distribution

We will begin our examination of the classical matching distribution by deriving its moments. As with many discrete distributions on the non-negative integers, it is easiest to examine the moments by first looking at the factorial moments and then using these to obtain the raw moments. Abramson (1973) has previously derived the factorial moments of the distribution and Feller (1968, p. 231) has derived the mean and variance of the distribution. We give a full derivation of all factorial and raw moments, the moment generating function, and the central moments up to fourth order. In Theorems 1-3 below we derive the form of the factorial moments and raw moments and the moment generating function. As with the general form of the distribution, these results generalise the case of a Poisson distribution with unit scale.

**THEOREM 1 (Factorial moments):** The factorial moments of the matching distribution are:
$$\mathbb{E}((K)_r) = \mathbb{I}(r \leq n).$$

**THEOREM 2 (Raw moments):** Let $S(r, i)$ denote the Stirling numbers of the second kind. The raw moments of the matching distribution are:
$$\mathbb{E}(K^r) = \sum_{i=0}^{\min(r,n)} S(r, i).$$
In the case where $r \leq n$ this reduces to the Bell numbers:
$$\mathbb{E}(K^r) = B_r.$$

**THEOREM 3 (MGF):** The moment generating function for the matching distribution is:

$$m_K(t) = \sum_{i=0}^{n} \frac{(e^t - 1)^i}{i!}.$$

The form of these moments is quite interesting in its own right. The factorial moments of the distribution are unit value for all $r \leq n$ and zero thereafter. In the special case where $n = \infty$ the distribution degenerates to the Poisson distribution with unit scale, which is known to have unit factorial moments and raw moments equal to the Bell numbers. Some important moments of the distribution are shown in Theorem 4 below.

**THEOREM 4 (Central moments):** Some important moments of the matching distribution are:

$$\mathbb{E}(K) = \begin{cases} 0 & \text{if } n = 0 \\ 1 & \text{if } n \geq 1 \end{cases} \qquad \mathbb{V}(K) = \begin{cases} 0 & \text{if } n \leq 1, \\ 1 & \text{if } n \geq 2. \end{cases}$$

$$\mathbb{Skew}(K) = \begin{cases} \text{NA} & \text{if } n \leq 1 \\ 0 & \text{if } n = 2 \\ 1 & \text{if } n \geq 3 \end{cases} \qquad \mathbb{Kurt}(K) = \begin{cases} \text{NA} & \text{if } n \leq 1, \\ 1 & \text{if } n = 2, \\ 3 & \text{if } n = 3, \\ 4 & \text{if } n \geq 4. \end{cases}$$

In the special cases where $n = 0$ or $n = 1$ the matching distribution is a point-mass on $K = 0$ or $K = 1$ respectively, and the moments for those cases reflect this. In the case where $n = 2$ the matching distribution is equiprobable on the values $K = 0$ and $K = 2$, yielding a symmetric and extreme platykurtic distribution (known to be the most platykurtic distribution). In the case where $n = 3$ the matching distribution is positively skewed and mesokurtic, and in the most general case where $n \geq 4$ the distribution is positively skewed and leptokurtic. The fixed values of the main central moments for large $n \geq 4$, and the slowly changing nature of the moment generating function, suggests that the distribution does not change much with respect to the size parameter $n$ once this value is already large. In particular, we see that the moment generating function has the form of a partial exponential generating function up to $n$, so in a neighbourhood of $t = 0$ the early terms are much larger than the later terms.

In Figure 1 we plot the distribution for size values $n = 1, \ldots, 12$ and confirm this fact. This shows that there is very little change in the distribution once $n > 6$. (At this point there is no discernible change in the barplot.) To confirm this observation, we plot the SSE comparing each distribution to the limiting case $n = \infty$ (i.e., the Poisson distribution with unit scale) in Figure 2.[2] From this latter plot we can confirm that the SSE diminishes rapidly, and for $n > 6$ it is no greater than $6 \times 10^{-6}$, which shows very little deviation from the limiting case. This latter result can be formalised with analytical bounds on the difference between probabilities in the matching distribution and probabilities in the Poisson distribution with unit scale (see e.g., DasGupta 2005, pp. 384-385). An interesting result is that there are alternating sign for these differences, though they diminish rapidly to zero (Ibid, p. 385).

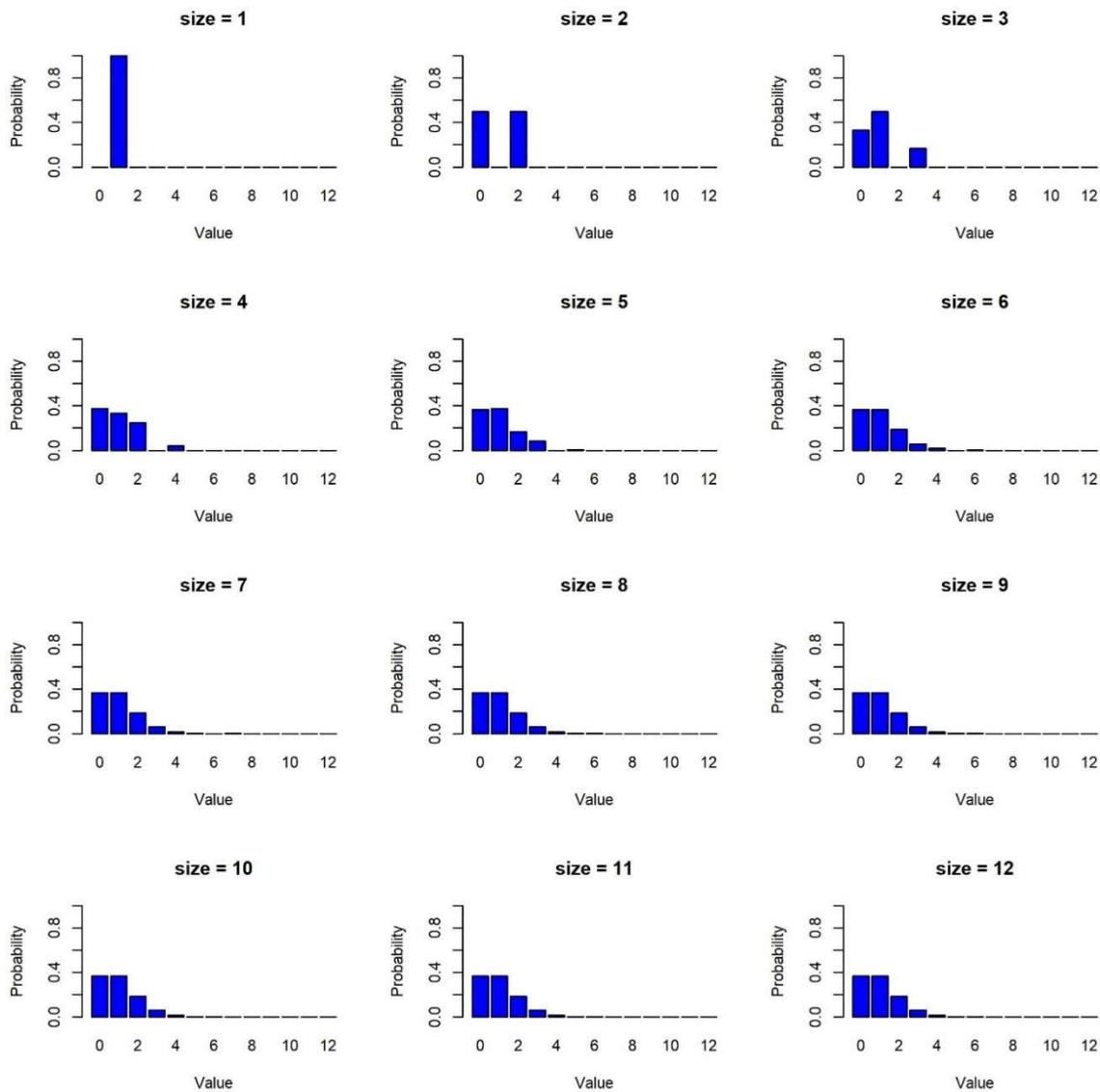

**FIGURE 1:** The classical matching distribution for sizes $n = 1, \ldots, 12$

---

[2] The SSE for each size value is defined as $\text{SSE}(n) \equiv \sum_{k=0}^{\infty}(\text{Match}(k|n) - \text{Match}(k|\infty))^2$. The computation for the figure truncates this sum at a finite upper bound, but this does not make any discernible difference in the plot.

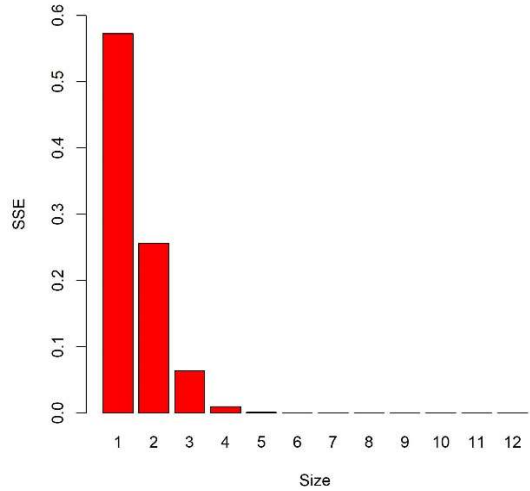

**FIGURE 2:** SSE for matching distributions of sizes $n = 1, \ldots, 12$ (compared to $n = \infty$)

The explicit form for the matching distribution involves a sum of rapidly decreasing numbers with alternating sign. This form is not particularly helpful for computational purposes, since sums of this kind commonly leads to problems of arithmetic underflow. In order to compute the mass function of the distribution, it is useful to characterise this function in recursive form. In Theorem 5 we give a recursive form for the probabilities in the distribution. In Theorem 6 we establish a recursive form comparing mass values at a fixed $k$ as $n$ increases. Both of these forms are simple consequences of the recursive properties of the subfactorial numbers.

**THEOREM 5 (Recursive form):** The mass function of the matching distribution satisfies the recursive equations:

$$\text{Match}(n|n) = \frac{1}{n!},$$

$$\text{Match}(k|n) = \frac{k+1}{n-k}[(n-k-1) \cdot \text{Match}(k+1|n) + (k+2) \cdot \text{Match}(k+2|n)].$$

**COROLLARY:** For a fixed size value $n$, let $\text{m}(k) \equiv \log \text{Match}(k|n)$ denote the log-probabilities for the matching distribution. These values satisfy the recursive equations:

$$\text{m}(n) = -\sum_{i=1}^{n} \log(i),$$

$$\text{m}(k) = \log(k+1) - \log(n-k) + \text{logsumexp}\begin{pmatrix} \log(n-k-1) + \text{m}(k+1), \\ \log(k+2) + \text{m}(k+2) \end{pmatrix}.$$

**THEOREM 6 (Recursive form):** The mass function of the matching distribution satisfies the recursive equation:

$$\text{Match}(k|n+1) = \frac{n-k}{n-k+1} \cdot \text{Match}(k|n) + \frac{k+1}{n-k+1} \cdot \text{Match}(k+1|n).$$

Consequently, for all fixed $k$ we have:

$$\lim_{n \to \infty} \frac{\text{Match}(k|n+1)}{\text{Match}(k|n)} = 1.$$

For computational purposes, it is useful to compute the distribution in log-space (i.e., compute the log-probabilities rather than the probabilities) and then convert back to regular probability space at the end of the computation. Consequently, the best way to compute the mass function is using the log-space recursive equations in the corollary to Theorem 5. The stability of the distribution for large $n$ is exhibited by the recursive equation and limiting result in Theorem 6. We defer computation of the mass function for now, since we will first want to generalise this distribution to handle a broader model for matching that allows results that are "better than random". The classical matching distribution gives the baseline behaviour of the matching number when allocation is done at random, and so it gives a null distribution for this hypothesis, which can serve as a basis for hypothesis testing for better allocation. In the next section we generalise the matching distribution to explicitly model allocation that is better than random.

## 3. Generalising the matching distribution

The essence of the classical problem of coincidences is that each selection of an item is assumed to occur via simple random sampling without replacement, such that all permutations over the set of items are equiprobable. This represents the premise that the allocation of items is based on purely random "guesses" or "draws" from the person doing the allocation. However, in certain kinds of matching problems, it is possible that the allocator may have some partial control over the matching. For example, consider the game where an allocator attempts to match baby photos to adults, where there is one baby photo for each adult. If the allocator is completely unable to match characteristics in the photos with characteristics of the adults then we would hypothesise that the selection is essentially just a random permutation of photos. However, if the allocator has some ability to match characteristics in the photos, he should tend to perform better than a random selection. Indeed, testing whether allocation is "better than random" is one of the primary purposes of the matching distribution.

Consideration of this kind of case leads to a desire to generalise the matching distribution to allow an additional parameter that will tend to increase the number of matches in some natural way. There are many ways that one could go about seeking this generalisation.[3] In this paper we will consider the matching mechanism as a two-step process. In the first step, for each item, suppose there is a fixed probability $\theta$ that the allocator will know the correct allocation for the item and allocate it accordingly. In the second step, all remaining items are allocated via a random permutation (i.e., via simple random sampling without replacement). This two-step process can be described mathematically by letting $L$ denote the (random) number of items with known allocation, so that the number of matches is then given by:

$$K_n^* = L + K_{n-L} \qquad L \sim \text{Bin}(n, \theta).$$

The resulting distribution of the matching number $K_n^*$ is then given by the convolution:

$$\text{Match}(k|n, \theta) = \sum_{\ell=0}^{k} \text{Bin}(\ell|n, \theta) \cdot \text{Match}(k - \ell|n - \ell)$$

$$= \sum_{\ell=0}^{k} \frac{n!}{(n-\ell)!\,\ell!} \cdot \theta^\ell (1-\theta)^{n-\ell} \cdot \frac{1}{(k-\ell)!} \cdot \frac{\mathcal{D}(n-k)}{(n-k)!}$$

$$= \frac{1}{k!} \cdot \frac{\mathcal{D}(n-k)}{(n-k)!} \sum_{\ell=0}^{k} \frac{n!}{(n-\ell)!} (1-\theta)^{n-k} \cdot \frac{k!}{(k-\ell)!\,\ell!} \cdot \theta^\ell (1-\theta)^{k-\ell}$$

$$= \text{Match}(k|n)\,(1-\theta)^{n-k} \sum_{\ell=0}^{k} \frac{n!}{(n-\ell)!} \text{Bin}(\ell|k, \theta).$$

We take this distribution to be a reasonable model of the matching process in cases where there may be some ability by the allocator to match the items "better than random".

**DEFINITION (Generalised matching distribution):** This is a family of discrete distributions characterised by the probability mass function:

$$\text{Match}(k|n, \theta) = \text{Match}(k|n)\,(1-\theta)^{n-k} \sum_{\ell=0}^{k} \frac{n!}{(n-\ell)!} \text{Bin}(\ell|k, \theta) \qquad k = 0, \ldots, n,$$

where we have a **size parameter** $n \in \mathbb{N}$ and probability parameter $0 \leq \theta \leq 1$. The generalised matching distribution reduces to the classical matching distribution when $\theta = 0$. □

---

[3] Another interesting generalisation, which we will not pursue in detail here, is to generalise using the distribution with moment generating function proportional to $\sum_{i=0}^{n} (\lambda(e^t - 1))^i / i!$, converging to the Poisson distribution as $n \to \infty$. Unfortunately, this only leads to a valid distribution when $0 \leq \lambda \leq 1$, which allows a lower number of matches but does not allow a larger number of matches. This is an interesting distribution, but it does not allow modelling of cases where allocation is "better than random" so it is unsuitable for our purposes.

Unfortunately, the mass function for the generalised distribution cannot be simplified further; it is a rather cumbersome form that presents some computational challenges. It is interesting to note that the mass function for the generalised matching distribution relates to the mass function for the classical case through the expectation of a falling factorial involving a binomial random variable. Specifically, we have $\text{Match}(k|n,\theta) = \text{Match}(k|n) \cdot (1-\theta)^{n-k} \cdot \Pi(k,\theta)$, where $\Pi(n,k,\theta) = \mathbb{E}((n)_B)$ for a random variable $B \sim \text{Bin}(k,\theta)$. The reader should note that the quantity $\Pi(n,k,\theta)$ is not a standard "factorial moment" for the binomial random variable, since those are of the form $\mathbb{E}((B)_r)$, not $\mathbb{E}((n)_B)$.

One effective way to compute the mass function for the generalised matching distribution is to first compute all the mass function values for all classical matching distributions up to the required size (using the recursive equations in Theorem 5) and then apply the convolution equation $\text{Match}(k|n,\theta) = \sum_{\ell=0}^{k} \text{Bin}(\ell|n,\theta) \cdot \text{Match}(k-\ell|n-\ell)$ to computation the mass values for the generalised distribution. It is advisable to conduct all these computations in log-space to avoid arithmetic underflow. Probability functions for the matching distribution are available in the **stat.extend** package in **R** (O'Neill and Fultz 2021). This package gives standard functions **dmatching**, **pmatching**, **qmatching** and **rmatching** respectively for the probability mass function, cumulative distribution function, quantile function, and random generation function for the distribution. (All the plots and computations in this paper were produced by computing the probabilities using these functions.)

In the special cases where $n=0$ or $n=1$ the generalised matching distribution is a point-mass on $K_n^* = 0$ or $K_n^* = 1$ respectively, just as in the classical case. For larger values the form of the distribution is more complicated, but it preserves the property that $\mathbb{P}(K_n^* = n-1) = 0$, reflecting the fact that it is impossible to match all but one item in any permutation. In the special case where $\theta = 0$ the generalised distribution reduces down to the classical form and in the special case where $\theta = 1$ it denegerates down to a point-mass distribution on $n$. Later we will show that the distribution is stochastically increasing in $\theta$, so these special cases give us the extremes. In Figure 3 we plot the distribution for size values $n = 1, \ldots, 12$ which shows some of these properties. Once $n$ becomes large the distribution converges to an asymptotic form for the sum of a binomial random variable and a Poisson random variable with unit scale, and of course, for large $n\theta$ the binomial part will tend to dominate the Poisson part, so the distribution will be close to a binomial distribution.

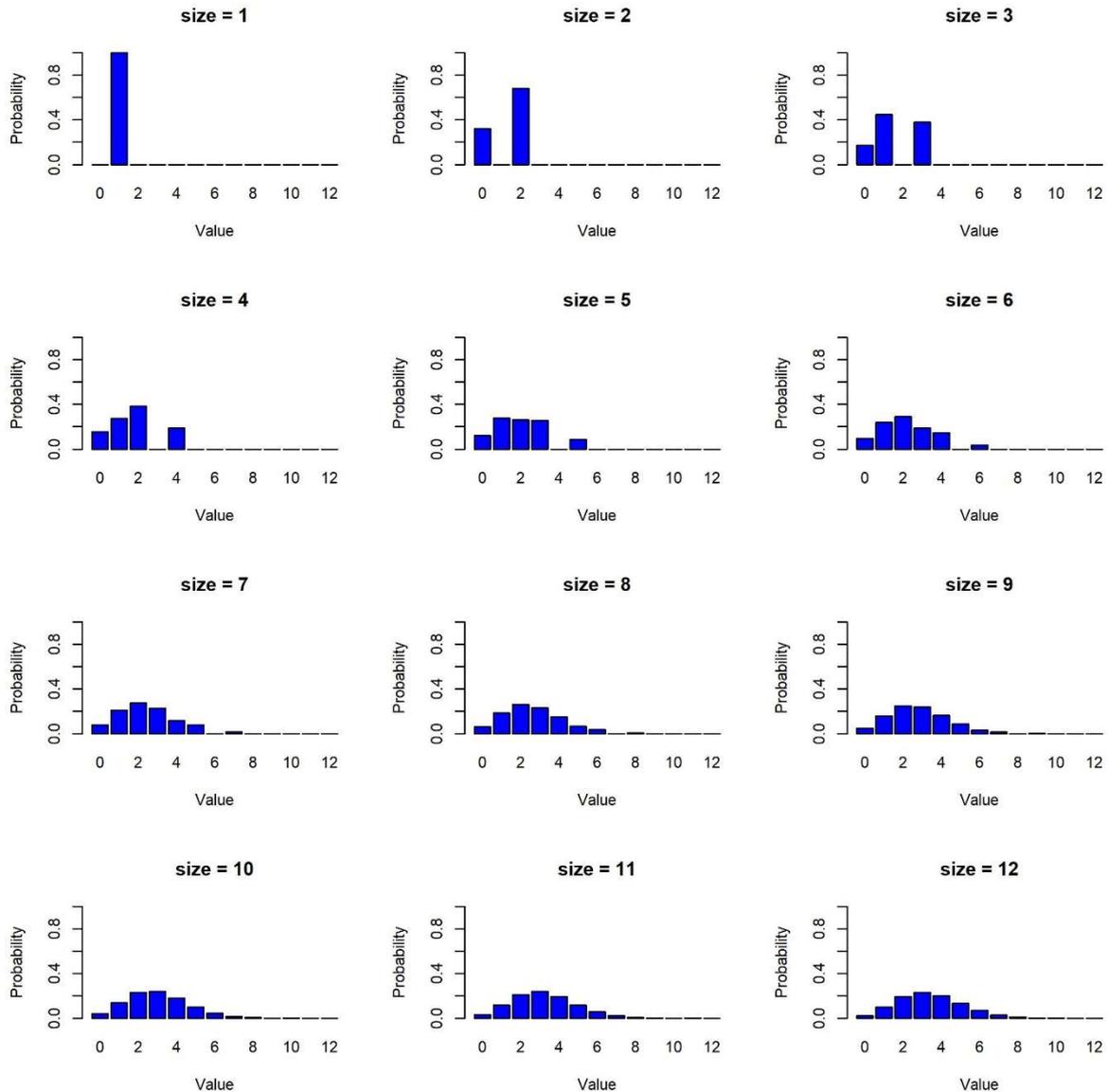

**FIGURE 3:** The generalised matching distribution for sizes $n = 1, \ldots, 12$ with $\theta = 0.2$

In Theorem 7 below we derive the moment generating function of the generalised matching distribution. It is possible to derive expressions for the moments using this expression or by applying the law of iterated expectations conditioning on the value *L*. The general forms for the factorial moments and raw moments are messy, and not particularly illuminating. It is also quite messy to obtain the higher-order moments of the distribution. In Theorem 8 we give the mean and variance of the distribution, which are polynomials of the probability parameter; the reader can easily verify that these moments reduce down to those of the classical matching distribution in the case where $\theta = 0$. In Theorem 9 we show that the generalised distribution obeys a central limit theorem as $n \to \infty$, so long as $0 < \theta < 1$.

**THEOREM 7 (MGF):** The moment generating function for the generalised distribution is:

$$m_{K_n^*}(t) = \sum_{i=0}^{n} \frac{(e^t - 1)^i}{i!} \sum_{\ell=0}^{n-i} \binom{n}{\ell} (\theta e^t)^\ell (1-\theta)^{n-\ell}.$$

**THEOREM 8 (Important central moments):** The mean, variance, skewness and kurtosis of the generalised distribution are given respectively by:

$$\mathbb{E}(K_n^*) = 1 + n\theta - \theta^n,$$

$$\mathbb{V}(K_n^*) = 1 - \theta^{2n} + n(\theta - \theta^2 - \theta^{n-1} - \theta^n + 2\theta^{n+1}),$$

$$\mathbb{Skew}(K_n^*) = \frac{\begin{pmatrix} 1 + n\theta(1 - 3\theta + 2\theta^2) \\ -\frac{n(n-1)}{2}\theta^{n-2} - n(2n-1)\theta^{n-1} \\ +\frac{5n^2 - 3n + 2}{2}\theta^n + 3n(n+1)\theta^{n+1} - 3n(n+1)\theta^{n+2} \\ -3n\theta^{2n-1} - 3n\theta^{2n} + 6n\theta^{2n+1} - 2\theta^{3n} \end{pmatrix}}{[1 - \theta^{2n} + n(\theta - \theta^2 - \theta^{n-1} - \theta^n + 2\theta^{n+1})]^{3/2}},$$

$$\mathbb{Kurt}(K_n^*) = 3 + \frac{\begin{pmatrix} 1 + n\theta - 7n\theta^2 + 12n\theta^3 - 6n\theta^4 \\ -\frac{n(n-1)(n-2)}{6}\theta^{n-3} - \frac{3n(n-1)^2}{2}\theta^{n-2} \\ -\frac{n(n^2 + 3n - 8)}{2}\theta^{n-1} - \frac{49n^3 - 84n^2 - 61n + 6}{6}\theta^n \\ -2n(2n^2 - 3n)\theta^{n+1} - 6n(n^2 + 3n + 2)\theta^{n+2} \\ +4n(n+1)(n+2)\theta^{n+3} - n(5n-2)\theta^{2n-2} \\ -2n(7n-2)\theta^{2n-1} - (n^2 - 12n - 2)\theta^{2n} \\ +12n(2n+1)\theta^{2n+1} - 12n(2n+1)\theta^{2n+2} \\ -12n\theta^{3n} + 24n\theta^{3n+1} - 6\theta^{4n} \end{pmatrix}}{[1 - \theta^{2n} + n(\theta - \theta^2 - \theta^{n-1} - \theta^n + 2\theta^{n+1})]^2}.$$

When $0 < \theta < 1$ and $n$ is large we obtain the asymptotic equivalences:

$$\mathbb{E}(K_n^*) \sim 1 + n\theta \qquad \mathbb{Skew}(K_n^*) \sim \frac{1 + n\theta(1-\theta)(1-2\theta)}{[1 + n\theta(1-\theta)]^{3/2}},$$

$$\mathbb{V}(K_n^*) \sim 1 + n\theta(1-\theta) \qquad \mathbb{Kurt}(K_n^*) \sim 3 + \frac{1 + n\theta(1-\theta)(6\theta^2 - 6\theta + 1)}{[1 + n\theta(1-\theta)]^2}.$$

This shows that the distribution is asymptotically unskewed and mesokurtic when $0 < \theta < 1$. (When $\theta = 0$ we have $\mathbb{Skew}(K_n^*) \sim 1$ and $\mathbb{Kurt}(K_n^*) \sim 4$ for the Poisson with unit rate.)

**THEOREM 9 (Central limit theorem):** If $0 < \theta < 1$ we get the following asymptotic result:

$$\mathbb{P}\left(\sqrt{n} \cdot \frac{K_n^*/n - 1/n - \theta}{\sqrt{1/n + \theta(1-\theta)}} \leq z\right) \to \Phi(z) \qquad \text{as } n \to \infty.$$

It is worth noting some intuition about the central limit theorem for the generalised distribution. From the fact that $K_n^* = L + K_{n-L}$, if $0 < \theta < 1$ we can see that when $n \to \infty$ the binomial random variable $L$ will come to dominate this sum and $K_n^*/L$ will converge to unity. Since the binomial distribution converges to the normal distribution (using the classical central limit theorem) this means that $K_n^*$ also converges to a normal random variable in this limit. The form of the quantity used in Theorem 8 is the standardised value of $K_n^*$ using its asymptotic mean and variance in Theorem 8.

The main value of our generalised matching distribution is that it allows us to model allocation that is "better than random" and is of increasing accuracy. The case where $\theta = 0$ corresponds to the classical matching distribution (where items under consideration are allocated at random) and the case where $\theta = 1$ gives a point-mass distribution on $K_n^* = n$. Between these extremes it can be shown that the parameter $\theta$ has a monotonic effect on the number of matches. To see this, we note that $\{K_n | n \in \mathbb{N}\}$ is a pure-birth process, which means that $K_n + 1 \geq K_{n+1}$ for all $n \in \mathbb{N}$. Consequently, for all values $0 \leq \ell < n$ we have:

$$\Delta(k, \ell) \equiv \mathbb{P}(K_n^* \leq k | L = \ell) - \mathbb{P}(K_n^* \leq k | L = \ell + 1)$$
$$= \mathbb{P}(L + K_{n-L} \leq k | L = \ell) - \mathbb{P}(L + K_{n-L} \leq k | L = \ell + 1)$$
$$= \mathbb{P}(\ell + K_{n-\ell} \leq k) - \mathbb{P}(\ell + K_{n-\ell-1} + 1 \leq k)$$
$$= \mathbb{P}(\ell + K_{n-\ell} \leq k) - \mathbb{P}(\ell + (K_{n-\ell-1} + 1) \leq k) \geq 0.$$

This establishes that $K_n^*$ is stochastically increasing in $L$, and since $L$ is stochastically increasing in $\theta$ it then follows that $K_n^*$ is stochastically increasing in $\theta$. If we again exclude the extremes and look at the case where $0 < \theta < 1$ the matching distribution obeys a central limit theorem shown in Theorem 9.

Since the mass function for the generalised matching distribution is a weighted sum of binomial mass functions, it is a polynomial in $\theta$. To facilitate this analysis, and some further analysis later in the paper, we will use the succinct notation $M(k, \ell) \equiv \text{Match}(k - \ell | n - \ell)$ to denote relevant values of the mass function for the classical matching distribution, which then gives:

$$\text{Match}(k|n, \theta) = \sum_{\ell=0}^{k} \text{Bin}(\ell|n, \theta) \cdot M(k, \ell) \qquad M(k, \ell) = \frac{1}{(k-\ell)!} \sum_{i=0}^{n-k} \frac{(-1)^i}{i!}.$$

This mixture form for the generalised matching distribution is useful for computation. We can use recursive methods to compute the classical matching distribution and then use the mixture result to generalise to the generalised matching distribution.

## 4. Computing the generalised matching distribution

As previously noted above, the explicit form for the mass function for the matching distribution involves a sum of terms with alternating sign. Such formulae are notoriously problematic for computation since they often lead to arithmetic underflow problems if one tries to compute via the explicit form. Moreover, it is best to compute probability functions in log-space to avoid arithmetic underflow from small probabilities. Consequently, the best way to compute the mass function for the classical matching distribution is work in log-space using the recursive form for the mass function (using the corollary to Theorem 5) and then convert back to regular probability space at the end of the computation. For the generalised matching distribution, we can compute the log-probabilities from the mass function using its convolution representation with the classical matching distribution and the binomial distribution. In Algorithm 1 below we implement this computational method, dealing separately with some trivial special cases.

```
                    ALGORITHM 1: Generalised Matching Distribution

Input:       Size parameter n (positive integer or Inf)
             Probability parameter θ (probability value)
             Logical value log (specifying whether output is log-probability)
Output:      Vector of probabilities/log-probabilities from the
             generalised matching distribution mass function for k = 0,…,n

######################## Deal with some special cases ########################

#Deal with special case where n = 1
#Distribution is a point-mass on one
if (n = 1) {
  LOGPROBS <- [-Inf, 0] }

#Deal with special case where 1 < n < Inf and θ = 1
#Distribution is a point-mass on n
if (1 < n < Inf)&(θ = 1) {
  LOGPROBS <- [-Inf, …, -Inf, 0] (vector has length = n+1) }

#Deal with special case where n = Inf and θ = 0
#Distribution is the Poisson distribution with unit rate
if (n = Inf)&(θ = 0) {
  LOGPROBS <- log(Pois(0:n|1)) }

#Deal with special case where n = Inf and θ > 0
#Distribution is a point-mass on infinity
if (n = Inf)&(θ > 0) {
  stop('Error: Distribution is a point-mass on infinity') }

########################### Deal with general case ##########################

#Deal with non-trivial case where 1 < n < Inf and θ < 1
if (1 < n < Inf)&(θ < 1) {
```

```
  #Compute the classical matching distribution by recursive method
  BASE.LOGPROBS <- Vector indexed by k = 0,…,n
                   (all starting values are -Inf)
  BASE.LOGPROBS[n] <- -logfactorial(n)
  for each i = 1,…,n {
    k <- n-i
    TERM1 <- log(n-k-1) + BASE.LOGPROBS[k+1]
    TERM2 <- ifelse(k < n-1, log(k+2) + BASE.LOGPROBS[k+2], -Inf)
    BASE.LOGPROBS[k] <- log(k+1) - log(n-k) + logsumexp(TERM1, TERM2) }
  BASE.LOGPROBS <- BASE.LOGPROBS - logsumexp(BASE.LOGPROBS)

  #Compute the matrix M
  M <- Matrix with rows indexed by k = 0,…,n and columns indexed by l = 0,…,n
       (all starting values are -Inf)
  for each k = 0,…,n {
  for each l = 0,…,k {
    M[k, l] <- BASE.LOGPROBS[n-l, k-l] } }

  #Compute the generalised matching distribution using mixture method
  LOGPROBS <- Vector indexed by k = 0,…,n
              (all starting values are -Inf)
  LOGBINOM <- log(Bin(0:n|n, θ)
  for each k = 0,…,n {
    LOGPROBS[k] <- logsumexp(LOGBINOM + M[k, ]) }
  LOGPROBS <- LOGPROBS - logsumexp(LOGPROBS) }

#Return output
if (log) { LOGPROBS } else { exp(LOGPROBS) }
```

The computation method in Algorithm 1 is stable and accurate and allows the user to compute the mass function for any allowable parameter inputs within the computational power of the machine.[4] The algorithm computes the probability mass function for the generalised matching distribution, but it also forms the basis for computing other probability functions, including the cumulative distribution function and quantile function. In the **stat.extend** package this algorithm is used in the functions **dmatching**, **pmatching** and **qmatching** for the mass function, cumulative distribution function and quantile function. The latter functions involve some further computational steps but use the same algorithm for computing the underlying log-probabilities for the distribution. Rather than producing the mass function (or other functions) over the full range $k = 0, ..., n$ these functions are modified to instead take an arbitrary input vector for the argument values and compute outputs over these argument values (i.e., they are "vectorised" versions of the probability mass function, cumulative distribution function and quantile function). (It is worth noting that Figures 1-3 above, showing the mass functions of the classical and generalised matching distribution, were each produced using the **dmatching** function to compute the probabilities.)

---

[4] Note that the parameter input $n = \infty$ and $\theta > 0$ leads to a point-mass distribution on infinity, which is not a proper distribution and cannot be represented in finite computation. The algorithm gives an error message in this case which note the true distribution.

Algorithm 1 can be extended to compute the highest-density region (HDR) for the generalised matching distribution by combining it with the discrete HDR algorithm presented in O'Neill (2021). It is simple to create other user-friendly functions that compute the moments of the distribution and other aspects of interest. The **stat.extend** package contains the function **HDR.matching** to compute the HDR from stipulated parameters for the distribution and a specified minimum coverage probability. It also contains the function **moments.matching** which computes the first four central moments of the generalised matching function.[5]

In the code below we generate and plot the probability functions for the generalised matching distribution with $n = 12$ and $\theta = 0.2$. The plots are shown in Figure 4 below. In addition to producing these probability plots we also compute the 95% HDR of the distribution. (Our commands are shown in blue and the output is in black.)

```
#Load library and set parameters
library(stat.extend)
n    <- 12
PROB <- 0.2

#Compute and plot the probability mass function
PDF <- dmatching(0:n, size = n, prob = PROB)
barplot(PDF, names.arg = 0:n, col = 'blue',
        xlab = 'Value', ylab = 'Probability')

#Compute and plot the cumulative distribution function
CDF <- pmatching(0:n, size = n, prob = PROB)
barplot(CDF, names.arg = 0:n, col = 'blue',
        xlab = 'Value', ylab = 'Cumulative Probability')

#Compute and plot the quantile function
PROBS <- (0:100)/100
QUANT <- qmatching(PROBS, size = n, prob = PROB)
plot(PROBS, QUANT, pch = 16,
     xlab = 'Probability', ylab = 'Quantile')

#Compute and plot pseudo-random values
set.seed(1)
RAND  <- rmatching(10^4, size = n, prob = PROB)
PROPS <- table(factor(RAND, levels = 0:n))/length(RAND)
barplot(PROPS, col = 'red', xlab = 'Value',
        ylab = 'Empirical Proportion')
```

---

[5] The function also has a logical option **include.sd** to specify whether the user wishes to include the standard deviation of the distribution in the output; by default this is not included.

```
#Compute the 95% HDR of the distribution
HDR.matching(cover.prob = 0.95, size = n, prob = PROB)

        Highest Density Region (HDR)

96.00% HDR for matching distribution with
12 trials and matching probability 0.2
Computed using discrete optimisation with
minimum coverage probability = 95.00%

1..7
```

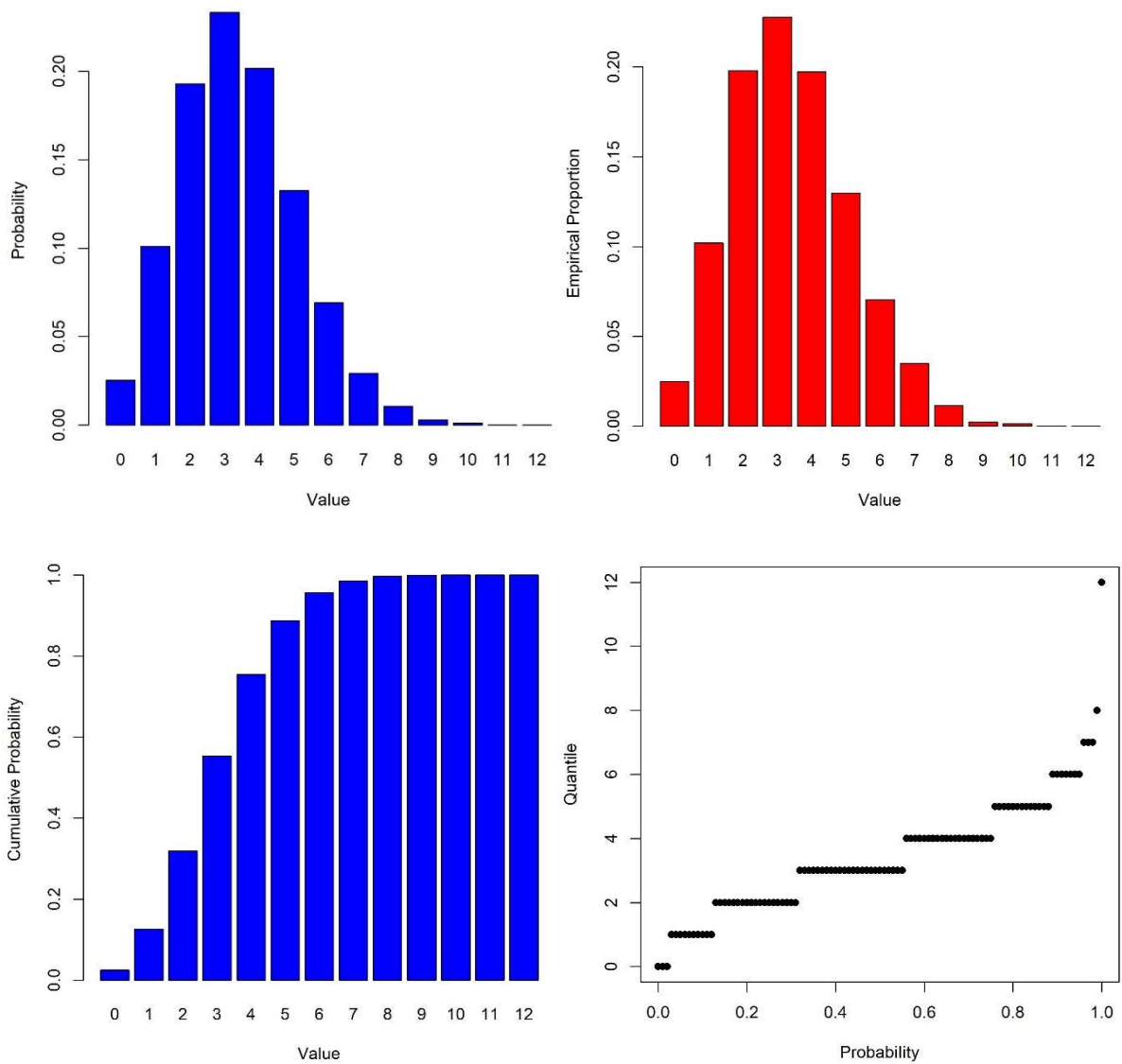

**FIGURE 4:** The generalised matching distribution with $n = 12$ and $\theta = 0.2$
**(top left)** probability density function; **(bottom left)** cumulative distribution function;
**(bottom right)** quantile function; **(top right)** proportions from random sample

Each of these probability functions other than the random-generation function is built on the computational method in Algorithm 1. For random generation from the generalised matching distribution we can use inverse-transformation sampling via quantile function, computed using the algorithm. Alternatively, we can generate values using direct simulation of the binomial matches and direct simulation of random permutations for the matching part. Interested readers can review code for the functions `dmatching`, `pmatching`, `qmatching` and `rmatching` in the `stat.extend` package to see how Algorithm 1 is adapted to create each probability function.

**5. Likelihood function and inference for the unknown probability parameter**

In applications of the matching problem, the size $n$ is fixed by the design of the matching game, but the probability parameter $\theta$ is unknown. In a matching game, this parameter represents the ability of the allocator to allocate the items "better than random" with a higher probability value representing greater likelihood of correct allocation. Here we will consider some methods to make inferences for this parameter. We will begin by considering some point-estimators and then later we will look at interval estimates. We will see that there are some essential coherence properties we would wish to impose on interval estimators for the parameter, which presents some challenges in finding an appropriate interval estimator.

In order to make inferences about the parameter $\theta$ we will derive the form of the log-likelihood function, score function, and information function, first for a single observation and then for a vector of IID observations. For purposes of numerical stability, and for other reasons, it is also useful to consider the problem framed in terms of the transformed parameter $\theta \mapsto \phi$ given by:

$$\theta = \frac{e^\phi}{e^\phi + e^{-\phi}} \qquad \phi = -\frac{1}{2} \cdot \log\left(\frac{1-\theta}{\theta}\right).$$

This transformation transforms the probability parameter $0 \leq \theta \leq 1$ to a value $\phi \in \overline{\mathbb{R}}$ on the extended real line, which means that subsequent steps use unconstrained optimisation instead of constrained optimisation. For purposes of succinct statement of the functions, we will use the notation $M(k, \ell) \equiv \text{Match}(k - \ell | n - \ell)$ introduced above. We begin by considering the case of a single observation $k$, corresponding to a single round of the matching game. The size parameter $n$ is fixed by the design of the matching game (i.e., it is a known constant). The log-likelihood function, score function and Hessian function are shown in Theorem 10 below, and the corresponding functions for the transformed parameter $\phi$ are shown in the corollary.

**THEOREM 10 (Score function and Hessian function):** For a single observation $k$ from the generalised matching distribution the log-likelihood function is:

$$\ell_k(\theta) = \log\left(\sum_{\ell=0}^{k} \text{Bin}(\ell|n,\theta) \cdot M(k,\ell)\right).$$

The corresponding score function and Hessian function are given respectively by:

$$s_k(\theta) = \frac{1}{\text{Match}(k|n,\theta)} \sum_{\ell=0}^{k} [\ell - n\theta] \cdot \frac{\text{Bin}(\ell|n,\theta)}{\theta(1-\theta)} \cdot M(k,\ell),$$

$$H_k(\theta) = \frac{1}{\text{Match}(k|n,\theta)} \sum_{\ell=0}^{k} \begin{bmatrix} (\ell-1)\ell \\ -2\ell(n-1)\theta \\ +n(n-1)\theta^2 \end{bmatrix} \cdot \frac{\text{Bin}(\ell|n,\theta)}{\theta^2(1-\theta)^2} \cdot M(k,\ell) - s_k(\theta)^2.$$

**COROLLARY:** Consider the transformed parameter $\theta \mapsto \phi$ given by:

$$\theta = \frac{e^\phi}{e^\phi + e^{-\phi}} \qquad \phi = -\frac{1}{2} \cdot \log\left(\frac{1-\theta}{\theta}\right).$$

The score function and Hessian function for $\phi$ are given (in terms of the parameter $\theta$) by:

$$s_k(\phi) = \frac{2}{\text{Match}(k|n,\theta)} \sum_{\ell=0}^{k} [\ell - n\theta] \cdot \text{Bin}(\ell|n,\theta) \cdot M(k,\ell),$$

$$H_k(\phi) = \frac{4}{\text{Match}(k|n,\theta)} \sum_{\ell=0}^{k} \begin{bmatrix} (\ell-1)\ell \\ -2\ell(n-1)\theta \\ +n(n-1)\theta^2 \end{bmatrix} \cdot \text{Bin}(\ell|n,\theta) \cdot M(k,\ell)$$
$$- 4(\theta - \tfrac{1}{2})s_k(\phi) - s_k(\phi)^2.$$

We can write the score function and Hessian function for $\theta$ in terms of the score function and Hessian function for $\phi$ as follows:

$$s_k(\theta) = \frac{s_k(\phi)}{2\theta(1-\theta)} \qquad H_k(\theta) = \frac{H_k(\phi) + 4(\theta - \tfrac{1}{2})s_k(\phi)}{4\theta^2(1-\theta)^2}.$$

We can generalise this likelihood analysis for the case where a single player plays a matching game with $n$ objects $m$ times, where we assume that these are IID trials with fixed parameters (i.e., we assume that the player is not getting better or worse at the game as they play it more). Suppose that we observe the outcomes $\boldsymbol{k} = (k_1, ..., k_m)$ over these games. Taking $n$ as fixed, the log-likelihood function, score function and Hessian function are then given by summing over the data points in the observed data vector.

It is simple to obtain the MLE for an IID sample from the generalised matching distribution by maximising the log-likelihood function using numerical methods. Since we have an explicit form for the Hessian function (and therefore the Fisher information) we can also compute the asymptotic standard error of the MLE and use this to obtain the standard asymptotic confidence interval for the unknown parameter. Alternatively, we can use bootstrapping to obtain a set of values for the MLE (based on resamples of the observed data **k**) and use this to form a bootstrap confidence interval for the unknown parameter. In either case, it is best to undertake both these computations in terms of the parameter $\phi$ and then transform back to obtain a corresponding MLE and confidence interval for $\theta$. Using the transformed parameter $\phi$ converts the problem to an unconstrained optimisation, which improves numerical stability, and also ensures that the resulting confidence interval respects the allowable parameter range.

The log-likelihood for IID data from the generalised matching distribution is unimodal, so there is a unique maximum likelihood estimator (MLE) for $\theta$. In the special case where $\bar{k}_m \leq 1$ the log-likelihood is monotonically decreasing in $\theta$, so the MLE is $\hat{\theta}_{\text{MLE}} = 0$. In the special case where $\bar{k}_m = n$ the log-likelihood is monotonically increasing in $\theta$, so the MLE is $\hat{\theta}_{\text{MLE}} = 1$. In the remaining cases the MLE is found using numerical methods. To compute the confidence interval, we recommend eschewing an equal-tail interval in favour of a "moving" interval. In view of the monotonicity properties of the log-likelihood, when $\bar{k}_n \leq 1$ the lower bound of the interval should be at zero and when $\bar{k}_n = n$ the upper bound of the interval should be at one. Consequently, at confidence level $1 - \alpha$ we recommend using the lower tail area:

$$\alpha_0 = \frac{\max(\bar{k}_n - 1, 0)}{n - 1}.$$

Use of this lower-tail area gives a "moving" interval which interpolates between these extremes to give a confidence interval that (roughly) takes account of the changing shape of the log-likelihood function. In particular, it should give the bounds required by the monotonicity properties of the log-likelihood function. In these cases, it is worth noting that the Hessian of the log-likelihood function vanishes at its extremes and so the standard asymptotic confidence interval performs poorly when the MLE for $\theta$ is near to zero or one (due to the failure of the underlying assumptions for that interval). Consequently, we recommend using the bootstrap interval in these cases.

The MLE and resulting confidence interval are implemented in the **MLE.match** function in the **stat.extend** package. This function allows the user to compute the confidence interval using the standard asymptotic form or via bootstrapping. In both cases the function uses the moving interval with the lower-tail area shown above. In the code below we give an example where we generate a random sample of size $m = 40$ from the matching distribution and use this to estimate the probability parameter. (Our commands are shown in blue and the output is shown in black.) In this example we use the bootstrap confidence interval, which gives better results in cases where the true probability parameter is near its boundaries (i.e., close to zero or one). As can be seen from the output, the MLE in this example is close to the true value of the probability parameter and the confidence interval contains the true value.

```
#Load library and set parameters
library(stat.extend)
n    <- 16
PROB <- 0.04

#Generate an IID random sample
set.seed(1)
DATA <- rmatching(40, size = n, prob = PROB)

#Generate the MLE and bootstrap confidence interval
MLE.matching(x = DATA, size = n,
             CI.method = 'bootstrap', conf.level = 0.99)

    Maximum likelihood estimator (MLE)

Data vector DATA containing 40 IID values
from the generalised matching
distribution with size = 16 and
unknown probability parameter

  MLE of probability parameter
     0.038502

  Maximum log-likelihood
     -67.96092

  Likelihood-per-data-point (maximised geometric mean)
     0.182862

  Bootstrap 99% CI for probability parameter
using 1000 resamples
     [0, 0.072687]
```

The MLE is just one way that the probability parameter can be estimated. A simpler estimator for the probability parameter can be obtained using the method of moments. Recalling from Theorem 8 that $\mathbb{E}(K_n^*) = 1 + n\theta - \theta^n$, and given the monotonicity properties of the likelihood function, the natural method-of-moments (MOM) estimator is given by the polynomial root:[6]

$$\hat{\theta}^n - n\hat{\theta} + \max(\bar{k}_n - 1, 0) = 0.$$

For $n = 1$ there is no valid MOM estimator since the matching distribution does not depend on $\theta$ (in this case any value for $\hat{\theta}$ satisfies the above equation). However, for $n > 1$ there is a unique MOM estimator, and it is simple to use root-finding methods to obtain this value.[7] It is easy to see that if $\bar{k}_n = 0$ or $\bar{k}_n = 1$ then $\hat{\theta} = 0$ and if $\bar{k}_n = n$ then $\hat{\theta} = 1$. Moreover, the estimate $\hat{\theta}$ is increasing in $\bar{k}_n$ for values between these two extremes. If $n$ is large we have the approximate MOM estimator:

$$\hat{\theta} \approx \frac{\max(\bar{k}_n - 1, 0)}{n - 1}.$$

In the code below we compute the MOM estimate and the approximate MOM estimator for the above example. As can be seen, in this example the MOM estimate is close to the MLE, but slightly further away from the true probability parameter than this latter estimate. Moreover, the approximate MOM estimate is still further from the true probability parameter.

```
#Set the MOM function
FUNC <- function(t) {
  t^n - n*t + max(mean(DATA)-1, 0) }

#Find the MOM estimator
uniroot(f = FUNC, lower = 0, upper = 1)$root

[1] 0.0390625

#Find the approximate MOM estimator
max(mean(DATA)-1, 0)/(n-1)

[1] 0.04666667
```

---

[6] Note that we have made a slight adjustment to the method of moments since $\mathbb{E}(K_n^*) \geq 1$ and so the observation $k = 0$ falls below the expected matches for any value of the probability parameter. In this special case, we estimate $\hat{\theta} = 0$ which is a sensible estimate given that the matching statistic is stochastically increasing in $\theta$.

[7] To see this, we observe that the MOM estimator is given by the roots $F(\hat{\theta}) = 0$ using the function:

$$F(\theta) = \theta^n - n\theta + \max(\bar{k}_n - 1, 0).$$

For $n > 1$ and $\theta < 1$ we have $F'(\theta) = -n(1 - \theta^{n-1}) < 0$ so $F$ is strictly decreasing. At the boundaries of the parameter range we have $F(0) = \max(\bar{k}_n - 1, 0) \geq 0$ and $F(1) = \max(\bar{k}_n, 1) - n \leq 0$. Consequently, under the condition that $n > 1$ there is a unique critical point $\hat{\theta}$ satisfying $F(\hat{\theta}) = 0$.

# 6. The matching test — hypothesis testing for the probability parameter

Instead of generating an interval estimate for the probability parameter it is sometimes useful to conduct a hypothesis test on this parameter. We will again consider the case where a player plays a matching game with $n$ objects $m$ times, giving rise to IID trials with fixed parameters, and we observe the outcomes $\boldsymbol{k} = (k_1, \ldots, k_m)$ over these games. We will consider both one-sided and two-sided tests using a simple null hypothesis positing a value $\theta_0$ for the probability parameter. This encompasses a range of useful tests for the matching probability.

Since the generalised matching distribution follows the monotone-likelihood ratio property, a reasonable test statistic for the hypothesis test is the mean number of matches over the trials, with a larger mean number of matches constituting evidence for a higher value of $\theta$ and a lower mean number of matches constituting evidence for a lower value of $\theta$.[8] This makes it useful to compute the distribution of the **total number of matches** $T_m = \sum_{i=1}^{m} K_i$ over $m$ trials where $K_1, \ldots, K_m \sim$ IID Match$(n, \theta)$. Using convolutions of the generalised matching distribution it is fairly simple to further generalise the distribution to add a **trials parameter** $m$ and return the distribution of the total number of matches under that number of trials. To this end, we will use the following notation to denote this more generalised distribution:

$$\text{Match}(t|n, m, \theta) = \mathbb{P}(T_m = t|n, m, \theta).$$

Of course, it is trivial to see that the special case where $m = 1$ gives the generalised matching distribution previously described, so we have $\text{Match}(t|n, 1, \theta) = \text{Match}(t|n, \theta)$. If $m$ is not too large, then we can compute the exact distribution for the total matches via convolutions of the generalised matching distribution. Contrarily, if $m$ is large, we can rely on the central limit theorem to approximate the generalised matching distribution with a normal distribution over the appropriate support.[9] We also note that there are some trivial special cases where we can perform exact computation easily, even when $m$ is large.[10]

---

[8] Another way to conduct the hypothesis test is to use the likelihood-ratio statistic for the generalised matching distribution. That method is more complex than the test we will use here.
[9] For the normal approximation we use the moments in Theorem 8 and we approximate the generalised distribution by $\text{Match}(t|n, m, \theta) \approx \text{N}(t|m\mathbb{E}(K_n^*), m\mathbb{V}(K_n^*))$ over $t = 0, \ldots, nm$. (Since it is impossible to get $n - 1$ matches in a single trial the support of this generalised distribution always excludes the point $t = mn - 1$.)
[10] Some trivial special cases are noted here. If $n = 0$ then the distribution is a point-mass on $t = 0$, if $n = 1$ then the distribution is a point-mass on $t = m$, if $\theta = 1$ then the distribution is a point-mass on $t = nm$. If $n = \infty$ and $\theta > 0$ then the distribution is a point-mass on $t = \infty$, and if $n = \infty$ and $\theta = 0$ then the distribution reduces to the Poisson distribution $\text{Match}(t|\infty, m, 0) = \text{Pois}(t|m)$. In these cases we would not use the normal approximation to the distribution even if $m$ is large.

The various functions for the generalised matching distribution in the **stat.extend** package actually do accommodate this generalisation. Each function allows an input for the number of trials $m$, with default behaviour setting $m = 1$ if the number of trials is not specified.[11] This distribution allows us to easily compute the p-value for any variation of the matching test. For the one-sided versions of the test (testing for parameter values above/below the null value) the p-value is the probability that the total number of matches $T_m$ would be no less than/no greater than the total observed matches $t_m$. For the two-sided version of the test (testing for parameter values different to the null value) the p-value is the probability that the total number of matches $T_m$ takes on a value no more probable than the total observed matches $t_m$. Each of these probabilities can easily be computed from the generalised matching distribution for $T_m$.

The canonical version of the matching test occurs when we test whether there is evidence that the matching done by a player is "better than random", which amounts to testing whether the probability parameter is zero. In this case we have the hypotheses:

$$H_0: \theta = 0 \qquad H_A: \theta > 0.$$

The matching test is implemented in the **matching.test** function in the **stat.extend** package. In the code below we implement the canonical matching test, and then a more specific one-sided matching test on the data vector previously generated from the generalised matching distribution. (By way of reminder, the vector **DATA** consists of $m = 40$ independent values from the generalised matching distribution with size $n = 16$ and probability $\theta = 0.04$.) As can be seen from the outcome of the two tests in this example, we find strong evidence that $\theta > 0$ (i.e., the matching is "better than random") but we find no evidence that $\theta > 0.05$.

```
#Perform a matching test for matching "better than random"
matching.test(DATA, size = 16)

        Matching test

data:   DATA
mean matches = 1.625, p-value = 0.0001726
alternative hypothesis: true prob is greater than 0
```

---

[11] This is done using the **trials** input in the functions. By default, this parameter is set to one, which gives the generalised matching distribution for a single trial. By default the function will compute the exact distribution using convolutions of the generalised matching distribution for a single trial. In the case where $m > 100$ it will switch (by default) to the normal approximation. The user can override this behaviour by using the **approx** input to tell the function whether or not to use the normal approximation to the distribution.

```
#Perform a matching test with a null probability of 5%
matching.test(DATA, size = 16, null.prob = 0.05)

        Matching test

data:   DATA
mean matches = 1.625, p-value = 0.8134
alternative hypothesis: true prob is greater than 0.05
```

Since we use the mean number of matches as our test statistic in the matching test, this test is a fairly straightforward inquiry into the distribution of the total number of matches in a set of trials of the matching game. It is natural to further inquire into the **power of this test**, to see how useful it is in identifying situations where we should reject a posited null value for the probability parameter in favour of some alternative hypothesis. We will examine the power of the canonical matching test over a range of values for the size $n$ and the trials $m$. When testing at significance level $\alpha$ the rejection-region for the canonical matching test is the set of values $t \geq t_*(\alpha)$ where the lower bound $t_*(\alpha)$ for the rejection-region is given by:[12]

$$t_*(\alpha) \equiv \min\{t = 0, \ldots, nm + 1 \mid \sum_{r=t}^{nm} \text{Match}(r|n, m, 0) < \alpha\}.$$

The power function for the canonical matching test is then given by:

$$\text{Power}(\theta|n, m, \alpha) = \mathbb{P}(T_m \geq t_*(\alpha)|n, m, \theta),$$

This power function gives the probability of rejecting the null hypothesis $H_0: \theta = 0$ given the true value of the probability parameter. Ideally, we would like the power to converge to one whenever $\theta > 0$ (i.e., whenever the alternative hypothesis is true).

In Figure 5 below we plot the power functions for a 5% significance level canonical matching test for the parameter combinations $n = 4, 6, 8, 10$ and $m = 1, 2, 3, 4, 5$. As can be seen, even for modest values for the size and trial parameters the power function increases quite rapidly with respect to the probability parameter $\theta$. We can also see that for a fixed value of $nm$ the test favours using a higher size parameter $n$ and a lower trial parameter $m$. The general shape of the power functions in Figure 1 illustrates convergence towards full power for values $\theta > 0$ (i.e., for all values in the alternative hypothesis) as the size and trials parameter increase. This is desirable behaviour for the test, and it reflects the fact that $T_m/nm \to \theta/m$ in probability under broad limiting conditions on either parameter.

---

[12] The special case where $t_*(\alpha) = nm + 1$ is a value outside the support of the distribution; this reflects the case where the rejection-region is empty. This will occur in some cases where the parameters are sufficiently small, or the significance level is sufficiently low, such that $\mathbb{P}(T_m = nm|n, m, 0) \geq \alpha$.

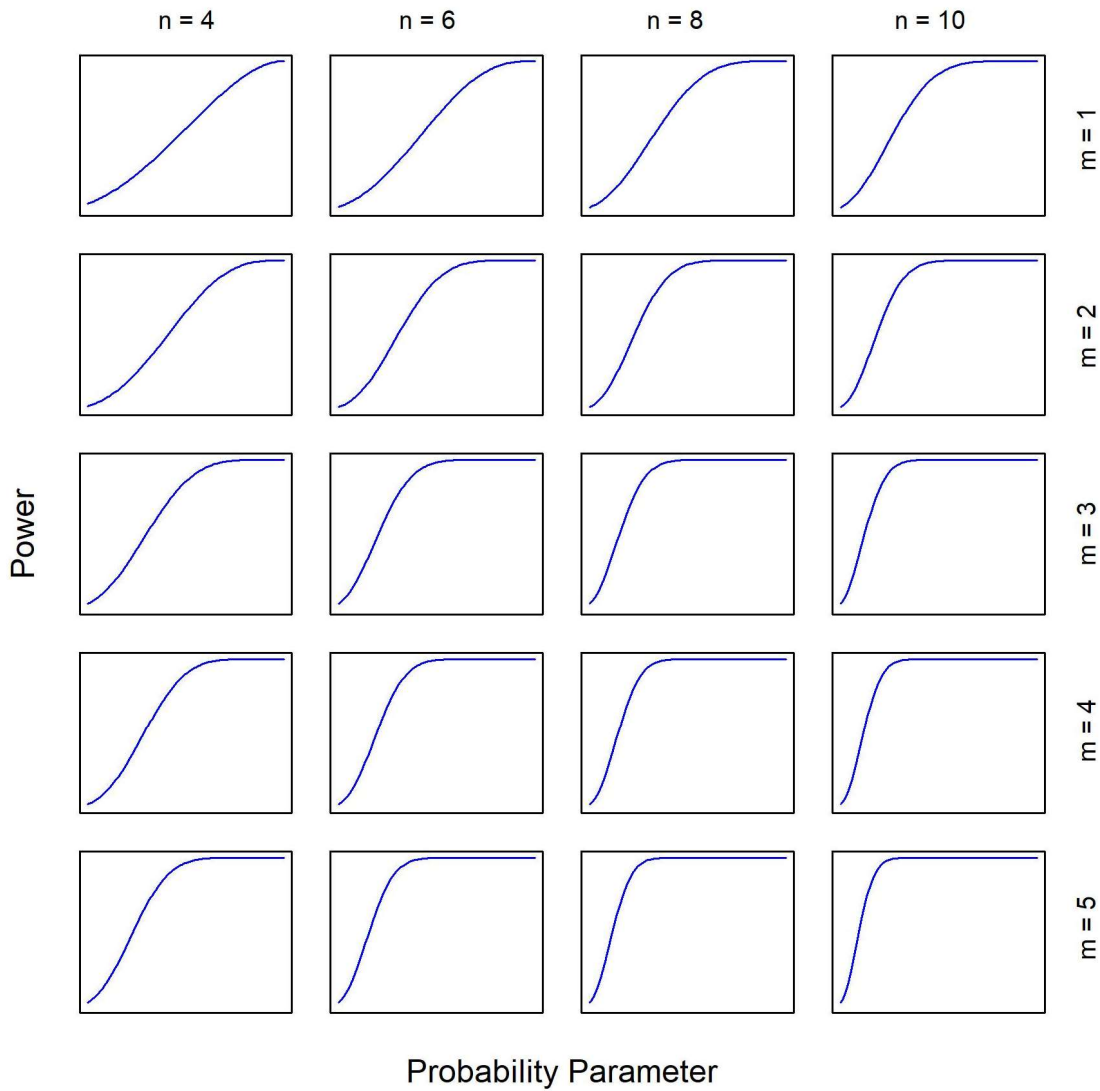

**FIGURE 5:** Power functions for canonical matching test at significance level $\alpha = 0.05$
(axis ranges are both probability values from zero to one)

It turns out that our power function will converge to this ideal power function if $n \to \infty$ or if $n > 1$ and $m \to \infty$. (In the case where $n = 0$ or $n = 1$ the generalised matching distribution is a point-mass distribution and the matching test has power zero. This holds even if we take the limit $m \to \infty$.) Before demonstrating this we will first show that the mean number of matches allows us to perfectly estimate a positive probability parameter under either of these limit conditions. Applying the moment results in Theorem 8 and rules for means and variances of sums of IID random variables, we get:

$$\mathbb{E}\left(\frac{T_m}{nm}\right) = \frac{1 + n\theta - \theta^n}{nm} \qquad \mathbb{V}\left(\frac{T_m}{nm}\right) = \frac{1 - \theta^{2n} + n(\theta - \theta^2 - \theta^{n-1} - \theta^n + 2\theta^{n+1})}{n^2 m}.$$

For $\theta > 0$ and under either of the stated limit conditions we have the asymptotic equivalence $\mathbb{E}(T_m/nm) \to \theta/m$ and $\mathbb{V}(T_m/nm) \to 0$. Consequently, the estimator $T_m/nm$ converges in probability to $\theta$ under either limit condition. This shows that $T_m/n$ is a consistent estimator for $\theta$ under either limit condition, but it also allows us to show that the power of the canonical matching test converges to one under the stated limiting conditions over all parameter values in the alternative hypothesis. Under either limiting condition we have $T_m/nm \to \theta/m$ which means that $t_*(\alpha)/nm \to 0$. Thus, for any value $\theta > 0$ we get:

$$\text{Power}(\theta|n, m, \alpha) = \mathbb{P}(T_m \geq t_*(\alpha)|n, m, \theta) = \mathbb{P}\left(\frac{T_m}{nm} \geq \frac{t_*(\alpha)}{nm} \bigg| n, m, \theta\right) \to 1.$$

One special case of the matching test that is noteworthy is the case where $n = 2$, such that the player in the matching game is trying to match a pair of objects in their correct order in each trial. In this case, in a single trial there are only two possible matches — the incorrect way of matching the objects, which gives $K = 0$, and the correct way of matching the objects, which gives $K = 2$. The player will *know* the correct order for the objects (not just guess it correctly via a random choice) if the player knows the place of at least one of the two objects, which occurs with probability $1 - (1 - \theta)^2 = 2\theta - \theta^2 = \theta(2 - \theta)$. If the player does not know the correct order for the objects then the player will still get the correct match with probability ½ due to randomly matching the two objects. Consequently, the probability of correctly matching both the objects in the matching game is:

$$\phi \equiv \mathbb{P}(K = 2) = (2\theta - \theta^2) + \tfrac{1}{2}(1 - 2\theta + \theta^2) = \frac{1 + 2\theta - \theta^2}{2},$$

and we have the simplified distributions:

$$\frac{K_1}{2}, \ldots, \frac{K_m}{2} \sim \text{IID Bern}(\phi) \qquad \frac{T_m}{2} \sim \text{Bin}(m, \phi).$$

It is simple to show that $\phi$ is a strictly increasing function of $0 \leq \theta \leq 1$, which means that the canonical matching distribution effectively reduces to an exact binomial test of the hypotheses:

$$H_0: \phi = \tfrac{1}{2} \qquad H_A: \phi > \tfrac{1}{2}.$$

In the code below we show an example of the canonical matching test for some data generated with $n = 2$ and we confirm that this gives the same p-value as the exact binomial test computed with the `binom.test` function in the `stats` package (R code team 2021).[13] (Note that the "probability of success" mentioned in the latter test is the parameter $\phi$, not the parameter $\theta$.)

---

[13] The p-values are identical to within a small tolerance which is due to rounding error in the computations.

```r
#Generate a new IID random sample with size = 2
set.seed(1)
NEWDATA <- rmatching(100, size = 2, prob = PROB)
table(NEWDATA)
```

```
NEWDATA
 0  2
51 49
```

```r
#Perform a matching test for matching "better than random"
(TEST1 <- matching.test(NEWDATA, size = 2))
```

```
        Matching test

data:  NEWDATA
mean matches = 0.98, p-value = 0.6178
alternative hypothesis: true prob is greater than 0
```

```r
#Perform an exact binomial test
SUCCESSES <- sum(NEWDATA)/2
TRIALS    <- length(NEWDATA)
(TEST2 <- binom.test(SUCCESSES, n = TRIALS,
                     alternative = 'greater'))
```

```
        Exact binomial test

data:  SUCCESSES and TRIALS
number of successes = 49, number of trials = 100,
p-value = 0.6178
alternative hypothesis: true probability
of success is greater than 0.5
95 percent confidence interval:
 0.403865 1.000000
sample estimates:
probability of success
                  0.49
```

```r
#Perform an exact binomial test
TEST1$p.value - TEST2$p.value
```

```
[1] 2.220446e-16
```

It is possible to form an alternative matching test using the likelihood-ratio-statistic instead of the mean/total number of matches. This has certain known advantages, including the fact that the Neyman-Pearson lemma ensures efficiency of the test. Nevertheless, the simpler form of test we offer here using the mean/total number of matches as the test statistic has the advantage of simplicity and maintains reasonable power even for modest values of the parameters.

## 7. Summary and conclusion

In this paper we have examined the properties of the classical matching distribution and a useful generalisation of this distribution. The classical matching distribution arises in the classical "problem of coincidences" — an antique problem dating back to the early eighteenth century. In this problem a player attempts to match the unknown order of a finite set of objects, and the order chosen by the player is taken to be a random permutation equivalent to simple-random-sampling without replacement. Our generalised distribution models the case where a player is first able to identify known matches for objects with a fixed probability (for each object), with the ordering of the remaining objects occurring "at random" via a random permutation. This generalisation allows us to deal with cases of matching games where the player has some ability to match the objects under consideration "better than random".

We have given a comprehensive analysis of the generalised matching distribution, including derivation of its probability mass function, moment generating function, main central moments and asymptotic behaviour. We have also developed a useful recursive algorithm to compute the probability mass function for the distribution and we have discussed how this algorithm can be extended to compute other probability functions. Finally, we have examined estimators, confidence intervals and hypothesis tests for the probability parameter in the distribution, under the condition that the size parameter is fixed by experimental design. All of this is implemented in functions in the **`stat.extend`** package for easy use by readers (see the table of available functions and their inputs below).

Our generalisation of the matching distribution allows a non-zero matching probability, so it can accommodate cases where matching is done "better than random". This allows us to use matching data to make inferences about the matching probability. In particular, we can use the canonical matching test to test for matching that is "better than random" and we can determine the power function for this test. Our view is that this is a useful and realistic generalisation of the classical "problem of coincidences". Indeed, matching analysis has occurred in contexts such as the hat-check problem and the secret-Santa problem, where the nature of the objects (or their owners/gives) may give some clues as to their proper order. In such cases it is useful to be able to model the possibility of a non-zero matching probability occurring prior to the random permutation of objects.

We hope that this paper piques the reader's interest in an interesting variation on an antique problem in probability theory. We also hope that it serves to assist analysis of this problem in a range of realistic cases. The generalised matching distribution we have examined in this paper is a natural extension of the classical matching distribution and it stands as an interesting and useful family of discrete distributions.

| Function | Inputs |
|---|---|
| `dmatching` | `x, size, trials = 1, prob = 0,`<br>`log = FALSE, approx = FALSE` |
| `pmatching` | `x, size, trials = 1, prob = 0,`<br>`lower.tail = TRUE, log.p = FALSE, approx = FALSE` |
| `qmatching` | `p, size, trials = 1, prob = 0,`<br>`lower.tail = TRUE, log.p = FALSE, approx = FALSE` |
| `rmatching` | `n, size, trials = 1, prob = 0` |
| `HDR.matching` | `cover.prob, size, trials = 1, prob = 0,`<br>`approx = FALSE` |
| `moments.matching` | `size, trials = 1, prob = 0, include.sd = FALSE` |
| `MLE.matching` | `size, prob = 0, CI.method = 'asymptotic',`<br>`conf.level = 0.95, bootstrap.sims = 10^3` |
| `matching.test` | `x, size, approx = FALSE,`<br>`null.prob = 0, alternative = 'greater',` |
| `lsubfactorial` | `x` |

**TABLE:** Functions for the generalised matching distribution
in the `stat.extend` package

# Appendix: Proof of Theorems

In this appendix we give proofs of the theorems in the main body of the paper. Many of the proofs involve the use of properties of the subfactorial numbers (see e.g., Hassani 2003 and Hassani 2004).

**PROOF OF THEOREM 1:** For all integer values $r > n$ we have $K - r + 1 \leq 0$ and so we get:

$$(K)_r = K(K-1)\ldots(K-r+1) = 0.$$

Contrarily, for the values $r = 0, \ldots, n$ we can use the change of variable $j = k - r$ to get:

$$\mathbb{E}((K)_r) = \sum_{k=0}^{n} (k)_r \cdot \text{Match}(k|n)$$

$$= \sum_{k=0}^{n} (k)_r \cdot \frac{1}{k!} \sum_{i=0}^{n-k} \frac{(-1)^i}{i!}$$

$$= \sum_{k=r}^{n} \frac{1}{(k-r)!} \sum_{i=0}^{n-k} \frac{(-1)^i}{i!}$$

$$= \sum_{j=0}^{n-r} \frac{1}{j!} \sum_{i=0}^{n-r-j} \frac{(-1)^i}{i!}$$

$$= \sum_{j=0}^{n-r} \text{Match}(j|n-r) = 1,$$

which establishes the result. ∎

**PROOF OF THEOREM 2:** The raw moments can be obtained from the factorial moments via:

$$\mathbb{E}(K^r) = \sum_{i=0}^{r} S(r,i) \cdot \mathbb{E}((K)_i),$$

where the values $S(r,i)$ are the Stirling numbers of the second kind (Moser and Wyman 1958; Rennie and Dobson 1969). Substituting the factorial moments from Theorem 1 gives:

$$\mathbb{E}(K^r) = \sum_{i=0}^{r} S(r,i) \cdot \mathbb{I}(i \leq n) = \sum_{i=0}^{\min(r,n)} S(r,i),$$

which was to be shown. The final part of the theorem follows directly from the fact that the Bell numbers are sums of Stirling numbers of the second kind. ∎

**PROOF OF THEOREM 3:** We will prove the stipulated form using substitution of the moments. Using the raw moments in Theorem 2 we have:

$$m_K(t) = 1 + \sum_{r=1}^{\infty} \frac{t^r}{r!} \cdot \mathbb{E}(K^r)$$

$$= \sum_{r=0}^{\infty} \frac{t^r}{r!} \sum_{i=0}^{r} S(r,i) \cdot \mathbb{I}(i \leq n)$$

$$= \sum_{r=0}^{\infty} \sum_{i=0}^{r} \frac{t^r}{r!} \cdot S(r,i) \cdot \mathbb{I}(i \leq n)$$

$$= \sum_{i=0}^{n} \sum_{r=i}^{\infty} \frac{t^r}{r!} \cdot S(r,i)$$

$$= \sum_{i=0}^{n} \frac{(e^t - 1)^i}{i!},$$

which was to be shown. ∎

**PROOF OF THEOREM 4:** To establish these results we will compute the central moments from the raw moments. The latter are given in Theorem 2 by the formula:

$$\mathbb{E}(K^r) = \sum_{i=0}^{\min(r,n)} S(r,i).$$

For $n = 0$ we have:

$$\mathbb{E}(K) = 0$$
$$\mathbb{E}(K^2) = 0 \qquad \mathbb{E}((K - \mu)^2) = 0,$$
$$\mathbb{E}(K^3) = 0 \qquad \mathbb{E}((K - \mu)^3) = 0,$$
$$\mathbb{E}(K^4) = 0 \qquad \mathbb{E}((K - \mu)^4) = 0.$$

For $n = 1$ we have:

$$\mathbb{E}(K) = 1$$
$$\mathbb{E}(K^2) = 1 \qquad \mathbb{E}((K - \mu)^2) = 0,$$
$$\mathbb{E}(K^3) = 1 \qquad \mathbb{E}((K - \mu)^3) = 0,$$
$$\mathbb{E}(K^4) = 1 \qquad \mathbb{E}((K - \mu)^4) = 0.$$

For $n = 2$ we have:

$$\mathbb{E}(K) = 1$$
$$\mathbb{E}(K^2) = 2 \qquad \mathbb{E}((K-\mu)^2) = 1,$$
$$\mathbb{E}(K^3) = 4 \qquad \mathbb{E}((K-\mu)^3) = 0,$$
$$\mathbb{E}(K^4) = 8 \qquad \mathbb{E}((K-\mu)^4) = 1.$$

For $n = 3$ we have:

$$\mathbb{E}(K) = 1$$
$$\mathbb{E}(K^2) = 2 \qquad \mathbb{E}((K-\mu)^2) = 1,$$
$$\mathbb{E}(K^3) = 5 \qquad \mathbb{E}((K-\mu)^3) = 1,$$
$$\mathbb{E}(K^4) = 14 \qquad \mathbb{E}((K-\mu)^4) = 3.$$

For $n \geq 4$ we have:

$$\mathbb{E}(K) = 1$$
$$\mathbb{E}(K^2) = 2 \qquad \mathbb{E}((K-\mu)^2) = 1,$$
$$\mathbb{E}(K^3) = 5 \qquad \mathbb{E}((K-\mu)^3) = 1,$$
$$\mathbb{E}(K^4) = 15 \qquad \mathbb{E}((K-\mu)^4) = 4.$$

The raw and central moments in the theorem follow directly from these values. ∎

**PROOF OF THEOREM 5:** The base equation in the theorem follows by substitution:

$$\text{Match}(n|n) = \frac{1}{(n-n)!\,n!} \cdot !(n-n) = \frac{1}{0!\,n!} \cdot !0 = \frac{1}{n!}.$$

To establish the recursive equation in the theorem we use the corresponding recursive equation for the subfactorial numbers (see e.g., Hassani 2004), which is:

$$!(n-k) = (n-k-1)[!(n-k-1) + !(n-k-2)].$$

Applying this equation gives:

$$\text{Match}(k|n) = \frac{1}{(n-k)!\,k!} \cdot !(n-k)$$

$$= \frac{1}{(n-k)!\,k!} \cdot (n-k-1)[!(n-k-1) + !(n-k-2)]$$

$$= \frac{k+1}{n-k} \cdot (n-k-1) \times \frac{1}{(n-k-1)!\,(k+1)!} \cdot !(n-k-1)$$

$$+ \frac{k+1}{n-k} \cdot (k+2) \times \frac{1}{(n-k-2)!\,(k+2)!} \cdot !(n-k-2)$$

$$= \frac{k+1}{n-k}[(n-k-1)\cdot \text{Match}(k+1|n) + (k+2)\cdot \text{Match}(k+2|n)],$$

which was to be shown. ∎

**PROOF OF THEOREM 6:** To establish the result we use the corresponding recursive equation for the subfactorial numbers, which is:

$$\mathcal{D}(x) = (x-1)[\mathcal{D}(x-1) + \mathcal{D}(x-2)].$$

Applying this equation with $x = n - k + 1$ gives:

$$\text{Match}(k|n+1) = \frac{1}{(n-k+1)!\,k!} \cdot \mathcal{D}(n-k+1)$$

$$= \frac{1}{(n-k+1)!\,k!} \cdot (n-k)[\mathcal{D}(n-k) + \mathcal{D}(n-k-1)]$$

$$= \frac{n-k}{n-k+1} \times \frac{1}{(n-k)!\,k!} \cdot \mathcal{D}(n-k)$$

$$+ \frac{k+1}{n-k+1} \times \frac{1}{(n-k-1)!\,(k+1)!} \cdot \mathcal{D}(n-k-1)$$

$$= \frac{n-k}{n-k+1} \times \text{Match}(k|n) + \frac{k+1}{n-k+1} \times \text{Match}(k+1|n),$$

which was to be shown. ∎

**PROOF THEOREM 7:** Using Theorem 3 and the relationship $K_n^* = L + K_{n-L}$ we have

$$m_{K_n^*}(t) \equiv \mathbb{E}(e^{tK_n^*}) = \mathbb{E}(e^{tL + tK_{n-L}})$$

$$= \sum_{\ell=0}^{n} \mathbb{E}(e^{tL + tK_{n-L}} | L = \ell) \cdot \text{Bin}(\ell|n,\theta)$$

$$= \sum_{\ell=0}^{n} e^{t\ell} \cdot \mathbb{E}(e^{tK_{n-\ell}}) \cdot \text{Bin}(\ell|n,\theta)$$

$$= \sum_{\ell=0}^{n} e^{t\ell} \cdot m_{K_{n-\ell}}(t) \cdot \text{Bin}(\ell|n,\theta)$$

$$= \sum_{\ell=0}^{n} e^{t\ell} \cdot \left( \sum_{i=0}^{n-\ell} \frac{(e^t - 1)^i}{i!} \right) \cdot \text{Bin}(\ell|n,\theta)$$

$$= \sum_{\ell=0}^{n} \sum_{i=0}^{n-\ell} e^{t\ell} \cdot \frac{(e^t - 1)^i}{i!} \cdot \text{Bin}(\ell|n,\theta)$$

$$= \sum_{\ell=0}^{n} \sum_{i=0}^{n-\ell} e^{t\ell} \cdot \frac{(e^t - 1)^i}{i!} \cdot \binom{n}{\ell} \theta^\ell (1-\theta)^{n-\ell}$$

$$= \sum_{\ell=0}^{n} \sum_{i=0}^{n-\ell} \frac{(e^t - 1)^i}{i!} \cdot \binom{n}{\ell} (\theta e^t)^\ell (1-\theta)^{n-\ell}$$

$$= \sum_{i=0}^{n} \frac{(e^t - 1)^i}{i!} \sum_{\ell=0}^{n-i} \binom{n}{\ell} (\theta e^t)^\ell (1-\theta)^{n-\ell},$$

which was to be shown.

**PROOF OF THEOREM 8:** From Theorem 2 we have the general rule:

$$\mathbb{E}(K_{n-\ell}^r) = \sum_{i=0}^{\min(r,n-\ell)} S(r,i) = \sum_{i=0}^{r} S(r,i) \cdot \mathbb{I}(i \leq n-\ell)$$

$$= \sum_{i=0}^{r} S(r,i) \cdot \mathbb{I}(\ell \leq n-i),$$

and for all $r > 0$ we have $S(r,0) = 0$ so we can then remove the first index in the sum to get:

$$\mathbb{E}(K_{n-\ell}^r) = \sum_{i=1}^{r} S(r,i) \cdot \mathbb{I}(\ell \leq n-i) \qquad \text{for } r > 0.$$

Applying this rule for $r = 1, \ldots, 4$ gives the specific results:

$$\mathbb{E}(K_{n-\ell}) = \mathbb{I}(\ell < n),$$

$$\mathbb{E}(K_{n-\ell}^2) = \mathbb{I}(\ell < n) + \mathbb{I}(\ell < n-1),$$

$$\mathbb{E}(K_{n-\ell}^3) = \mathbb{I}(\ell < n) + 3 \cdot \mathbb{I}(\ell < n-1) + \mathbb{I}(\ell < n-2),$$

$$\mathbb{E}(K_{n-\ell}^4) = \mathbb{I}(\ell < n) + 7 \cdot \mathbb{I}(\ell < n-1) + 6 \cdot \mathbb{I}(\ell < n-2) + \mathbb{I}(\ell < n-3).$$

We will first use these results to obtain the first four raw moments. Using the law of iterated expectation, we have:

$$\mathbb{E}(K_n^*) = \mathbb{E}(L + K_{n-L}) = \mathbb{E}(\mathbb{E}(L + K_{n-L}|L))$$

$$= \sum_{\ell=0}^{n} \mathbb{E}(L + K_{n-L}|L = \ell) \cdot \text{Bin}(\ell|n,\theta)$$

$$= \sum_{\ell=0}^{n} \mathbb{E}(\ell + K_{n-\ell}) \cdot \text{Bin}(\ell|n,\theta)$$

$$= \sum_{\ell=0}^{n} (\ell + \mathbb{I}(\ell < n)) \cdot \text{Bin}(\ell|n,\theta)$$

$$= 1 + n\theta - \theta^n,$$

$$\mathbb{E}(K_n^{*2}) = \mathbb{E}((L + K_{n-L})^2) = \mathbb{E}(\mathbb{E}((L + K_{n-L})^2|L))$$

$$= \sum_{\ell=0}^{n} \mathbb{E}((L + K_{n-L})^2|L = \ell) \cdot \text{Bin}(\ell|n,\theta)$$

$$= \sum_{\ell=0}^{n} \mathbb{E}((\ell + K_{n-\ell})^2) \cdot \text{Bin}(\ell|n,\theta)$$

$$= \sum_{\ell=0}^{n} \mathbb{E}(\ell^2 + 2\ell K_{n-\ell} + K_{n-\ell}^2) \cdot \text{Bin}(\ell|n,\theta)$$

$$= \sum_{\ell=0}^{n} \binom{\ell^2 + 2\ell \mathbb{I}(\ell < n)}{+\mathbb{I}(\ell < n) + \mathbb{I}(\ell < n-1)} \cdot \text{Bin}(\ell|n,\theta)$$

$$= \left[ \begin{array}{c} n\theta(1-\theta+n\theta) + (2n\theta - 2n\theta^n) \\ +(1-\theta^n) + (1 - n(1-\theta)\theta^{n-1} - \theta^n) \end{array} \right]$$

$$= 2 + 3n\theta + n(n-1)\theta^2 - n\theta^{n-1} - (n+2)\theta^n.$$

$$\mathbb{E}(K_n^{*3}) = \mathbb{E}((L + K_{n-L})^3) = \mathbb{E}(\mathbb{E}((L + K_{n-L})^3|L))$$

$$= \sum_{\ell=0}^{n} \mathbb{E}((L + K_{n-L})^3|L = \ell) \cdot \text{Bin}(\ell|n,\theta)$$

$$= \sum_{\ell=0}^{n} \mathbb{E}((\ell + K_{n-L})^3) \cdot \text{Bin}(\ell|n,\theta)$$

$$= \sum_{\ell=0}^{n} \mathbb{E}(\ell^3 + 3\ell^2 K_{n-\ell} + 3\ell K_{n-\ell}^2 + K_{n-\ell}^3) \cdot \text{Bin}(\ell|n,\theta)$$

$$= \sum_{\ell=0}^{n} \left( \begin{array}{c} \ell^3 + 3\ell^2 \mathbb{I}(\ell < n) \\ +3\ell \mathbb{I}(\ell < n) + 3\ell \mathbb{I}(\ell < n-1) \\ +\mathbb{I}(\ell < n) + 3\mathbb{I}(\ell < n-1) + \mathbb{I}(\ell < n-2) \end{array} \right) \cdot \text{Bin}(\ell|n,\theta)$$

$$= \sum_{\ell=0}^{n} \left( \begin{array}{c} \ell^3 + 3\ell^2 + 6\ell + 5 \\ -(3n^2 + 6n + 5)\mathbb{I}(\ell = n) \\ -(3n+1)\mathbb{I}(\ell = n-1) \\ -\mathbb{I}(\ell = n-2) \end{array} \right) \cdot \text{Bin}(\ell|n,\theta)$$

$$= \left[ \begin{array}{c} n\theta(1 - 3\theta + 3n\theta + 2\theta^2 - 3n\theta^2 + n^2\theta^2) \\ +3n\theta(1 - \theta + n\theta) + 6n\theta + 5 \\ -(3n^2 + 6n + 5)\theta^n \\ -(3n+1)n(1-\theta)\theta^{n-1} \\ -\dfrac{n(n-1)}{2}(1-\theta)^2 \theta^{n-2} \end{array} \right]$$

$$= \left[ \begin{array}{c} 5 + n\theta(10 - 6\theta + 6n\theta + 2\theta^2 - 3n\theta^2 + n^2\theta^2) \\ -\dfrac{n(n-1)}{2}\theta^{n-2} - 2n(n+1)\theta^{n-1} - \left(5(n+1) + \dfrac{n(n-1)}{2}\right)\theta^n \end{array} \right],$$

$$\mathbb{E}(K_n^{*4}) = \mathbb{E}((L + K_{n-L})^4) = \mathbb{E}(\mathbb{E}((L + K_{n-L})^4|L))$$

$$= \sum_{\ell=0}^{n} \mathbb{E}((L + K_{n-L})^4|L = \ell) \cdot \text{Bin}(\ell|n,\theta)$$

$$= \sum_{\ell=0}^{n} \mathbb{E}((\ell + K_{n-L})^4) \cdot \text{Bin}(\ell|n,\theta)$$

$$= \sum_{\ell=0}^{n} \mathbb{E}(\ell^4 + 4\ell^3 K_{n-\ell} + 6\ell^2 K_{n-\ell}^2 + 4\ell K_{n-\ell}^3 + K_{n-\ell}^4) \cdot \text{Bin}(\ell|n,\theta)$$

$$= \sum_{\ell=0}^{n} \begin{pmatrix} \ell^4 + 4\ell^3 \mathbb{I}(\ell < n) \\ +6\ell^2 \mathbb{I}(\ell < n) + 6\ell^2 \mathbb{I}(\ell < n-1) \\ +4\ell \mathbb{I}(\ell < n) + 12\ell \mathbb{I}(\ell < n-1) + 4\ell \mathbb{I}(\ell < n-2) \\ \mathbb{I}(\ell < n) + 7 \cdot \mathbb{I}(\ell < n-1) \\ +6 \cdot \mathbb{I}(\ell < n-2) + \mathbb{I}(\ell < n-3) \end{pmatrix} \cdot \text{Bin}(\ell|n,\theta)$$

$$= \sum_{\ell=0}^{n} \begin{pmatrix} \ell^4 + 4\ell^3 + 12\ell^2 + 20\ell + 15 \\ -(4n^3 + 12n^2 + 20n + 15)\mathbb{I}(\ell = n) \\ -(6n^2 + 4n + 4)\mathbb{I}(\ell = n-1) \\ -(4n-1)\mathbb{I}(\ell = n-2) \\ -\mathbb{I}(\ell = n-3) \end{pmatrix} \cdot \text{Bin}(\ell|n,\theta)$$

$$= \begin{bmatrix} n\theta \begin{pmatrix} 1 - 7\theta + 7n\theta + 12\theta^2 - 18n\theta^2 \\ +6n^2\theta^2 - 6\theta^3 + 11n\theta^3 - 6n^2\theta^3 + n^3\theta^3 \end{pmatrix} \\ +4n\theta(1 - 3\theta + 3n\theta + 2\theta^2 - 3n\theta^2 + n^2\theta^2) \\ +12n\theta(1 - \theta + n\theta) + 20n\theta + 15 \\ -(4n^3 + 12n^2 + 20n + 15)\theta^n - (6n^2 + 4n + 4)n(1-\theta)\theta^{n-1} \\ -(4n-1)\dfrac{n(n-1)}{2}(1-\theta)^2\theta^{n-2} - \dfrac{n(n-1)(n-2)}{6}(1-\theta)^3\theta^{n-3} \end{bmatrix}$$

$$= \begin{bmatrix} 15 + n\theta \begin{pmatrix} 37 - 31\theta + 31n\theta + 20\theta^2 - 30n\theta^2 \\ +10n^2\theta^2 - 6\theta^3 + 11n\theta^3 - 6n^2\theta^3 + n^3\theta^3 \end{pmatrix} \\ -(4n^3 + 12n^2 + 20n + 15)\theta^n \\ -(6n^2 + 4n + 4)n(1-\theta)\theta^{n-1} \\ -(4n-1)\dfrac{n(n-1)}{2}(1-\theta)^2\theta^{n-2} \\ -\dfrac{n(n-1)(n-2)}{6}(1-\theta)^3\theta^{n-3} \end{bmatrix}$$

$$= \begin{bmatrix} 15 + n\theta \begin{pmatrix} 37 - 31\theta + 31n\theta + 20\theta^2 - 30n\theta^2 \\ +10n^2\theta^2 - 6\theta^3 + 11n\theta^3 - 6n^2\theta^3 + n^3\theta^3 \end{pmatrix} \\ -\dfrac{n(n-1)(n-2)}{6}\theta^{n-3} \\ +\dfrac{n(n-1)(n-2)}{2}\theta^{n-2} - (4n-1)\dfrac{n(n-1)}{2}\theta^{n-2} \\ +(4n-1)n(n-1)\theta^{n-1} - (6n^2 + 4n + 4)n\theta^{n-1} - \dfrac{n(n-1)(n-2)}{2}\theta^{n-1} \\ +(6n^2 + 4n + 4)n\theta^n - (4n^3 + 12n^2 + 20n + 15)\theta^n \\ +\dfrac{n(n-1)(n-2)}{6}\theta^n - (4n-1)\dfrac{n(n-1)}{2}\theta^n \end{bmatrix}$$

$$= \begin{bmatrix} 15 + n\theta \begin{pmatrix} 37 - 31\theta + 31n\theta + 20\theta^2 - 30n\theta^2 \\ +10n^2\theta^2 - 6\theta^3 + 11n\theta^3 - 6n^2\theta^3 + n^3\theta^3 \end{pmatrix} \\ -\dfrac{n(n-1)(n-2)}{6}\theta^{n-3} \\ -\dfrac{n(n-1)(3n+1)}{2}\theta^{n-2} \\ -\dfrac{n(5n^2+15n+8)}{2}\theta^{n-1} \\ +\dfrac{n^3 - 36n^2 - 97n - 90}{6}\theta^n \end{bmatrix}.$$

From these raw moments we can obtain the relevant central moments, which give the moments in the theorem. With a bit of algebra it can be shown that:

$$\mathbb{E}(K_n^*)^2 = \begin{pmatrix} 1 + 2n\theta + n^2\theta^2 \\ -2\theta^n - 2n\theta^{n+1} + \theta^{2n} \end{pmatrix},$$

$$\mathbb{E}(K_n^*)^3 = \begin{pmatrix} 1 + 3n\theta + 3n^2\theta^2 + n^3\theta^3 \\ -3\theta^n - 6n\theta^{n+1} - 3n^2\theta^{n+2} \\ +3\theta^{2n} + 3n\theta^{2n+1} - \theta^{3n} \end{pmatrix},$$

$$\mathbb{E}(K_n^*)^4 = \begin{pmatrix} 1 + 4n\theta + 6n^2\theta^2 + 4n^3\theta^3 + n^4\theta^4 \\ -4\theta^n - 12n\theta^{n+1} - 12n^2\theta^{n+2} - 4n^3\theta^{n+3} \\ +6\theta^{2n} + 12n\theta^{2n+1} + 6n^2\theta^{2n+2} \\ -4\theta^{3n} - 4n\theta^{3n+1} + \theta^{4n} \end{pmatrix}.$$

The variance of the distribution is:

$$\mathbb{V}(K_n^*) = \mathbb{E}(K_n^{*2}) - \mathbb{E}(K_n^*)^2$$
$$= (2 + 3n\theta + n(n-1)\theta^2 - n\theta^{n-1} - (n+2)\theta^n) - (1 + n\theta - \theta^n)^2$$
$$= (2 + 3n\theta + n(n-1)\theta^2 - n\theta^{n-1} - (n+2)\theta^n) - \begin{pmatrix} 1 + 2n\theta + n^2\theta^2 \\ -2\theta^n - 2n\theta^{n+1} + \theta^{2n} \end{pmatrix}$$
$$= 1 - \theta^{2n} + n(\theta - \theta^2 - \theta^{n-1} - \theta^n + 2\theta^{n+1}).$$

With a substantial amount of algebra we can establish the third and fourth central moments:

$$\mathbb{E}((K_n^* - \mathbb{E}(K_n^*))^3) = \mathbb{E}(K_n^{*3}) - 3\mathbb{E}(K_n^{*2})\mathbb{E}(K_n^*) + 2\mathbb{E}(K_n^*)^3$$
$$= \left[ 5 + n\theta(10 - 6\theta + 6n\theta + 2\theta^2 - 3n\theta^2 + n^2\theta^2) \right.$$
$$\left. -\dfrac{n(n-1)}{2}\theta^{n-2} - 2n(n+1)\theta^{n-1} - \left(5(n+1) + \dfrac{n(n-1)}{2}\right)\theta^n \right]$$
$$-3(2 + 3n\theta + n(n-1)\theta^2 - n\theta^{n-1} - (n+2)\theta^n)(1 + n\theta - \theta^n)$$
$$+2(1 + n\theta - \theta^n)^3$$

$$= \begin{pmatrix} 1 + n\theta(1 - 3\theta + 2\theta^2) \\ -\dfrac{n(n-1)}{2}\theta^{n-2} - n(2n-1)\theta^{n-1} \\ +\dfrac{5n^2 - 3n + 2}{2}\theta^n + 3n(n+1)\theta^{n+1} - 3n(n+1)\theta^{n+2} \\ -3n\theta^{2n-1} - 3n\theta^{2n} + 6n\theta^{2n+1} - 2\theta^{3n} \end{pmatrix},$$

$$\mathbb{E}((K_n^* - \mathbb{E}(K_n^*))^4) = \mathbb{E}(K_n^{*4}) - 4\mathbb{E}(K_n^{*3})\mathbb{E}(K_n^*) + 6\mathbb{E}(K_n^{*2})\mathbb{E}(K_n^*)^2 - 3\mathbb{E}(K_n^*)^4$$

$$= \begin{bmatrix} 15 + n\theta\begin{pmatrix} 37 - 31\theta + 31n\theta + 20\theta^2 - 30n\theta^2 \\ +10n^2\theta^2 - 6\theta^3 + 11n\theta^3 - 6n^2\theta^3 + n^3\theta^3 \end{pmatrix} \\ -\dfrac{n(n-1)(n-2)}{6}\theta^{n-3} \\ -\dfrac{n(n-1)(3n+1)}{2}\theta^{n-2} \\ -\dfrac{n(5n^2 + 15n + 8)}{2}\theta^{n-1} \\ +\dfrac{n^3 - 36n^2 - 97n - 90}{6}\theta^n \end{bmatrix}$$

$$-4\begin{bmatrix} 5 + n\theta(10 - 6\theta + 6n\theta + 2\theta^2 - 3n\theta^2 + n^2\theta^2) \\ -\dfrac{n(n-1)}{2}\theta^{n-2} - 2n(n+1)\theta^{n-1} \\ -\left(5(n+1) + \dfrac{n(n-1)}{2}\right)\theta^n \end{bmatrix}(1 + n\theta - \theta^n)$$

$$+6\begin{pmatrix} 2 + 3n\theta + n(n-1)\theta^2 \\ -n\theta^{n-1} - (n+2)\theta^n \end{pmatrix}(1 + n\theta - \theta^n)^2 - 3(1 + n\theta - \theta^n)^4$$

$$= \dfrac{1}{6}\begin{pmatrix} 24 + 42n\theta + 6n(3n-13)\theta^2 \\ -36n(n-2)\theta^3 + 18n(n-2)\theta^4 \\ -n(n^2 - 3n + 2)\theta^{n-3} - 9n(n-1)^2\theta^{n-2} \\ -3n(n^2 + 3n + 4)\theta^{n-1} - (49n^3 - 48n^2 - 25n + 6)\theta^n \\ -12n(2n^2 - 3n - 6)\theta^{n+1} - 36n(n^2 + 2)\theta^{n+2} \\ +24n(n^2 + 2)\theta^{n+3} - 12n(n-1)\theta^{2n-2} \\ -24n(2n-1)\theta^{2n-1} - (60n^2 - 36n - 12)\theta^{2n} \\ +36n(2n+1)\theta^{2n+1} - 36n(2n+1)\theta^{2n+2} \\ +36n\theta^{3n-1} - 36n\theta^{3n} + 72n\theta^{3n+1} - 18\theta^{4n} \end{pmatrix}.$$

(For brevity, we have omitted a substantial number of algebraic steps in this computation. The reader is invited to undertake the required algebra if they wish to do so. It is not particularly illuminating work.) We then have:

$$\mathbb{S}\text{kew}(K_n^*) = \dfrac{\begin{pmatrix} 1 + n\theta(1 - 3\theta + 2\theta^2) \\ -\dfrac{n(n-1)}{2}\theta^{n-2} - n(2n-1)\theta^{n-1} \\ +\dfrac{5n^2 - 3n + 2}{2}\theta^n + 3n(n+1)\theta^{n+1} - 3n(n+1)\theta^{n+2} \\ -3n\theta^{2n-1} - 3n\theta^{2n} + 6n\theta^{2n+1} - 2\theta^{3n} \end{pmatrix}}{[1 - \theta^{2n} + n(\theta - \theta^2 - \theta^{n-1} - \theta^n + 2\theta^{n+1})]^{3/2}},$$

$$\mathbb{Kurt}(K_n^*) = \frac{1}{6} \cdot \frac{\begin{pmatrix} 24 + 42n\theta + 6n(3n-13)\theta^2 \\ -36n(n-2)\theta^3 + 18n(n-2)\theta^4 \\ -n(n-1)(n-2)\theta^{n-3} - 9n(n-1)^2\theta^{n-2} \\ -3n(n^2+3n+4)\theta^{n-1} - (49n^3 - 48n^2 - 25n + 6)\theta^n \\ -12n(2n^2 - 3n - 6)\theta^{n+1} - 36n(n^2+2)\theta^{n+2} \\ +24n(n^2+2)\theta^{n+3} - 12n(n-1)\theta^{2n-2} \\ -24n(2n-1)\theta^{2n-1} - (60n^2 - 36n - 12)\theta^{2n} \\ +36n(2n+1)\theta^{2n+1} - 36n(2n+1)\theta^{2n+2} \\ +36n\theta^{3n-1} - 36n\theta^{3n} + 72n\theta^{3n+1} - 18\theta^{4n} \end{pmatrix}}{[1 - \theta^{2n} + n(\theta - \theta^2 - \theta^{n-1} - \theta^n + 2\theta^{n+1})]^2}$$

$$= 3 + \frac{\begin{pmatrix} 1 + n\theta - 7n\theta^2 + 12n\theta^3 - 6n\theta^4 \\ -\frac{n(n-1)(n-2)}{6}\theta^{n-3} - \frac{3n(n-1)^2}{2}\theta^{n-2} \\ -\frac{n(n^2+3n-8)}{2}\theta^{n-1} - \frac{49n^3 - 84n^2 - 61n + 6}{6}\theta^n \\ -2n(2n^2 - 3n)\theta^{n+1} - 6n(n^2+3n+2)\theta^{n+2} \\ +4n(n+1)(n+2)\theta^{n+3} - n(5n-2)\theta^{2n-2} \\ -2n(7n-2)\theta^{2n-1} - (n^2 - 12n - 2)\theta^{2n} \\ +12n(2n+1)\theta^{2n+1} - 12n(2n+1)\theta^{2n+2} \\ -12n\theta^{3n} + 24n\theta^{3n+1} - 6\theta^{4n} \end{pmatrix}}{[1 - \theta^{2n} + n(\theta - \theta^2 - \theta^{n-1} - \theta^n + 2\theta^{n+1})]^2}.$$

This establishes the moment results in the theorem. The asymptotic results are obtained by removing all higher-order terms of the form $\theta^{n+k}$ and then simplifying the expressions with additional algebra. ∎

**PROOF OF THEOREM 9:** Using the asymptotic equivalence results in Theorem 7 we have:

$$\frac{K_n^* - \mathbb{E}(K_n^*)}{\sqrt{\mathbb{V}(K_n^*)}} \sim \frac{1}{\sqrt{n}} \cdot \frac{K_n^* - 1 - n\theta}{\sqrt{\theta(1-\theta)}} = \sqrt{n} \cdot \frac{K_n^*/n - 1/n - \theta}{\sqrt{\theta(1-\theta)}}.$$

Since $K_n^*/n = L/n + K_{n-L}/n$ and $K_{n-L}/n \to 1/n \to 0$ in probability, we can apply Slutsky's theorem to obtain the asymptotic equivalence $K_n^*/n \sim (L+1)/n$. (Indeed, the asymptotic mean and variance shown in Theorem 8 matches the mean and variance of $L + 1$.) Since $L$ is a binomial random variable it is composed of IID Bernoulli values and the classical central limit theorem applies to $L/n$. Combining this asymptotic result with Slutsky's theorem gives the result to be shown. ∎

**PROOF OF THEOREM 10:** The log-likelihood function in the theorem follows immediately by taking the logarithm of the mass function for the generalised matching distribution. To obtain the score function and information function we will use the first and second-order logarithmic derivatives, which are:

$$\frac{d}{d\theta}\log p(\theta) = \frac{p'(\theta)}{p(\theta)},$$

$$\frac{d^2}{d\theta^2}\log p(\theta) = \frac{p''(\theta)}{p(\theta)} - \left(\frac{d}{d\theta}\log p(\theta)\right)^2.$$

With some algebraic manipulation, we can establish that:

$$\begin{aligned}
\frac{d}{d\theta}\text{Bin}(\ell|n,\theta) &= \frac{n!}{\ell!\,(n-\ell)!}\frac{d}{d\theta}\theta^\ell(1-\theta)^{n-\ell} \\
&= \frac{n!}{\ell!\,(n-\ell)!}\left[\ell\theta^{\ell-1}(1-\theta)^{n-\ell} - (n-\ell)\theta^\ell(1-\theta)^{n-\ell-1}\right] \\
&= \frac{n!}{\ell!\,(n-\ell)!}\theta^\ell(1-\theta)^{n-\ell}\left(\frac{\ell}{\theta} - \frac{n-\ell}{1-\theta}\right) \\
&= \frac{n!}{\ell!\,(n-\ell)!}\theta^\ell(1-\theta)^{n-\ell}\cdot\frac{\ell(1-\theta) - (n-\ell)\theta}{\theta(1-\theta)} \\
&= \frac{n!}{\ell!\,(n-\ell)!}\theta^\ell(1-\theta)^{n-\ell}\cdot\frac{\ell - n\theta}{\theta(1-\theta)} \\
&= [\ell - n\theta]\cdot\frac{\text{Bin}(\ell|n,\theta)}{\theta(1-\theta)},
\end{aligned}$$

$$\begin{aligned}
\frac{d^2}{d\theta^2}\text{Bin}(\ell|n,\theta) &= \frac{d}{d\theta}\left(\text{Bin}(\ell|n,\theta)\cdot\frac{\ell - n\theta}{\theta(1-\theta)}\right) \\
&= \frac{\ell - n\theta}{\theta(1-\theta)}\cdot\frac{d}{d\theta}\text{Bin}(\ell|n,\theta) + \text{Bin}(\ell|n,\theta)\cdot\frac{d}{d\theta}\frac{\ell-n\theta}{\theta(1-\theta)} \\
&= \frac{\ell - n\theta}{\theta(1-\theta)}\cdot\frac{d}{d\theta}\text{Bin}(\ell|n,\theta) \\
&\quad + \text{Bin}(\ell|n,\theta)\cdot\frac{-n\theta(1-\theta) - (\ell-n\theta)(1-\theta) + (\ell-n\theta)\theta}{\theta^2(1-\theta)^2} \\
&= \frac{\ell - n\theta}{\theta(1-\theta)}\cdot\frac{d}{d\theta}\text{Bin}(\ell|n,\theta) + \text{Bin}(\ell|n,\theta)\cdot\frac{2\ell\theta - \ell - n\theta^2}{\theta^2(1-\theta)^2} \\
&= [(\ell - n\theta)^2 + (2\ell\theta - \ell - n\theta^2)]\frac{\text{Bin}(\ell|n,\theta)}{\theta^2(1-\theta)^2} \\
&= [(\ell^2 - 2n\ell\theta + n^2\theta^2) + (2\ell\theta - \ell - n\theta^2)]\frac{\text{Bin}(\ell|n,\theta)}{\theta^2(1-\theta)^2} \\
&= [\ell(\ell-1) - 2\ell(n-1)\theta + n(n-1)\theta^2]\frac{\text{Bin}(\ell|n,\theta)}{\theta^2(1-\theta)^2}.
\end{aligned}$$

Substituting these derivatives into the forms for the logarithmic derivatives gives the stated score function and Hessian function. The corollary is easily obtained using the transformation:

$$\theta = \frac{e^{2\phi}}{e^{2\phi}+1} \qquad \phi = -\frac{1}{2}\cdot\log\left(\frac{1-\theta}{\theta}\right),$$

with corresponding derivatives (written in terms of $\theta$) given by:

$$\frac{d\theta}{d\phi} = 2\theta(1-\theta) \qquad \frac{d^2\theta}{d\phi^2} = 8(½-\theta)\theta(1-\theta),$$

$$\frac{d\phi}{d\theta} = \frac{1}{2\theta(1-\theta)} \qquad \frac{d^2\phi}{d\theta^2} = -\frac{½-\theta}{\theta^2(1-\theta)^2}.$$

Using the standard rules for transformations involving the gradient vector and Hessian matrix leads to the results stated in the corollary. ∎